\documentclass[prb,twocolumn,amssymb,amsmath,floatfix,showpacs]{revtex4}
\usepackage{bm,dcolumn,graphicx,hyperref}
\newcommand{\vect}[1]{\mathbf{#1}}
\newcommand{\kp}{\ensuremath{\vect{k} \cdot \vect{p}}}
\newcolumntype{w}[1]{D{.}{.}{#1}}

\begin{document}
\title{Valence-band mixing in first-principles envelope-function theory}
\author{Bradley A. Foreman}
\email{phbaf@ust.hk}
\affiliation{Department of Physics,
             Hong Kong University of Science and Technology,
             Clear Water Bay, Kowloon, Hong~Kong, China}

\begin{abstract}
This paper presents a numerical implementation of a first-principles
envelope-function theory derived recently by the author [B. A.
Foreman, Phys.\ Rev.\ B \textbf{72}, 165345 (2005)].  The examples
studied deal with the valence subband structure of GaAs/AlAs,
GaAs/Al$_{0.2}$Ga$_{0.8}$As, and In$_{0.53}$Ga$_{0.47}$As/InP (001)
superlattices calculated using the local density approximation to
density-functional theory and norm-conserving pseudopotentials without
spin-orbit coupling.  The heterostructure Hamiltonian is approximated
using quadratic response theory, with the heterostructure treated as a
perturbation of a bulk reference crystal.  The valence subband
structure is reproduced accurately over a wide energy range by a
multiband envelope-function Hamiltonian with linear renormalization of
the momentum and mass parameters.  Good results are also obtained over
a more limited energy range from a single-band model with quadratic
renormalization.  The effective kinetic-energy operator ordering
derived here is more complicated than in many previous studies,
consisting in general of a linear combination of all possible operator
orderings.  In some cases the valence-band Rashba coupling differs
significantly from the bulk magnetic Luttinger parameter.  The
splitting of the quasidegenerate ground state of no-common-atom
superlattices has non-negligible contributions from both short-range
interface mixing and long-range dipole terms in the quadratic density
response.
\end{abstract}


\maketitle

\section{Introduction}

In a recent paper, the author has developed a first-principles
multiband envelope-function theory for semiconductor
heterostructures. \cite{Fore05b} This theory can in principle provide
accurate numerical predictions of envelope-function Hamiltonian
parameters, but only if a reliable quasiparticle self-energy is used
as input.  On the other hand, if the input potential is taken from a
simple density-functional calculation, \cite{Fore07a} the numerical
values are less accurate, but the theory can still provide deep
insight into the basic physics of the interface and clarify various
limitations of commonly used envelope-function models.  The purpose of
the present article is to explore these topics using numerical
examples taken from density-functional calculations of the valence
subband structure in semiconductor superlattices.

In order to obtain the most direct comparison with envelope-function
theory as it is commonly used, the Hamiltonian was derived in Ref.\
\onlinecite{Fore05b} in the form of a matrix containing differential
operators and energy-independent functions of position.  Using a
differential equation of low order to describe an abrupt
heterojunction might seem at first glance to be a gross theoretical
blunder, since it is well known that bulk effective-mass theory
\cite{LuKo55} is valid only for slowly varying perturbing potentials.
Yet effective-mass theory is widely accepted as a valid lowest-order
approximation for the shallow impurity problem,
\cite{Kohn57b,Bass74,Pant78} even though the atomic impurity potential
is not slowly varying.  Its validity in this context is established by
the existence of rigorous methods for extending effective-mass theory
to include higher-order perturbations, \cite{Bass74,Pant78,Sham66}
including the rapidly varying part of the impurity
potential. \cite{BirPik74_sec27} These produce short-range and
long-range corrections to the effective Hamiltonian (each with its own
independent parameters), which lead to the well known chemical shift
and splitting of effective-mass degeneracies,
\cite{Bass74,Pant78,BirPik74_sec27} thereby bringing the rough
estimates of effective-mass theory into much closer agreement with
experiment.

A heterostructure is nothing but an assembly of atoms.  It is
therefore difficult to imagine a valid argument for accepting the use
of generalized effective-mass differential equations in the shallow
impurity problem, yet categorically rejecting them for small atomic
perturbations in heterostructures.  According to linear response
theory, the perturbation generated by a collection of small atomic
perturbations is to leading order just the superposition of the
individual perturbations.  The rapidly varying part of the
heterostructure potential \cite{Burt92,TakVol99b} can thus be handled
by the same techniques used in the impurity
problem. \cite{BirPik74_sec27} The validity of low-order differential
equations for shallow heterostructures---as for shallow
impurities---consequently rests on two fundamental assumptions:
\cite{LuKo55,BirPik74_sec27} that the envelope functions satisfying
these equations are slowly varying \cite{Burt92} (i.e., have a Fourier
transform limited to small wave vectors) within the energy range of
experimental interest, and that the atomic perturbations are truly
``shallow'' in real heterostructures.

The first assumption is confirmed by numerical examples in later
sections of this paper, in agreement with prior empirical
pseudopotential studies. \cite{Xia89,WaFrZu97,WaZu97,WaZu99} For the
second assumption, the definition of ``shallow'' is a relative one
that depends on the energy separation between the states included
explicitly in the envelope-function model and the remote states
treated as perturbations [see, e.g., Eq.\ (II.33) of Ref.\
\onlinecite{LuKo55}].  Consider the case of a GaAs/AlAs
heterostructure, where the valence-band offset is about 0.5 eV and the
energy gap is about 1.5 eV\@.  Treating GaAs/AlAs as a shallow
perturbation of the virtual crystal Al$_{0.5}$Ga$_{0.5}$As would be
expected to yield marginal results (at best) in a single-band
effective-mass model for the degenerate $\Gamma$ valence states.  Such
a model would be expected to give good results only for weaker
perturbations, such as those in a GaAs/Al$_{0.2}$Ga$_{0.8}$As
heterostructure (within the virtual-crystal approximation).  On the
other hand, treating GaAs/AlAs as a shallow perturbation would be
expected to produce good results in a multiband envelope-function
model that includes a few of the nearest conduction-band states
explicitly, since the remote bands are then more than 5 eV away from
the valence-band maximum.  (This method of making moderately ``deep''
perturbations ``shallow'' by including more bands in the Hamiltonian
was proposed by Keldysh for the deep impurity problem in bulk
semiconductors.\cite{Keld64}) All of the above expectations are
confirmed below by numerical examples for GaAs/Al$_{0.2}$Ga$_{0.8}$As,
GaAs/AlAs, and In$_{0.53}$Ga$_{0.47}$As/InP\@.

To be more specific, the multiband envelope-function Hamiltonian of
Ref.\ \onlinecite{Fore05b} was derived by treating the heterostructure
as a perturbation (within the pseudopotential approximation) of a
virtual bulk reference crystal.  The self-consistent potential energy
of the heterostructure was approximated using the linear and quadratic
terms of nonlinear response theory (see the preceding paper
\cite{Fore07a} for further details).  A finite-order envelope-function
Hamiltonian was constructed by using Luttinger--Kohn perturbation
theory \cite{LuKo55,BirPik74_sec27,Leib75,Leib77} to eliminate the
\kp\ and potential-energy coupling to remote bands, working to order
$k^4$ in the bulk reference Hamiltonian, \cite{TakVol99b} to order
$k^2$ in the linear response terms, \cite{Leib75,Leib77} and to order
$k^0$ in the quadratic response. \cite{TakVol99b} This theory is shown
here to work well in a 3-state model for the $\Gamma_{15\mathrm{v}}$
valence states (i.e., neglecting spin-orbit coupling) of a
GaAs/Al$_{0.2}$Ga$_{0.8}$As superlattice and in a 7-state $\{
\Gamma_{15\mathrm{c}}, \Gamma_{1\mathrm{c}}, \Gamma_{15\mathrm{v}} \}$
model for GaAs/AlAs and In$_{0.53}$Ga$_{0.47}$As/InP superlattices.
Examples illustrating the success of a 4-state $\{
\Gamma_{1\mathrm{c}}, \Gamma_{15\mathrm{v}} \}$ model for
In$_{0.53}$Ga$_{0.47}$As/InP superlattices have been given
elsewhere. \cite{Fore07c}

To test the limits of the single-band ($\Gamma_{15\mathrm{v}}$) model,
this paper also extends the theory of Ref.\ \onlinecite{Fore05b} to
include terms of order $\theta^2 k$ and $\theta^2 k^2$, where $\theta$
denotes the heterostructure perturbation.  This is shown to yield much
better predictions of the position-dependent effective masses of both
GaAs/AlAs and In$_{0.53}$Ga$_{0.47}$As/InP than the $O(\theta^1 k^2)$
theory of Ref.\ \onlinecite{Fore05b}.  The resulting superlattice band
structures are also of reasonably good accuracy, but over a more
limited energy range than the multiband models.  The principal
limitation of the single-band theory seems to be the presence of
spurious solutions generated by $O(\theta^0 k^4)$ terms in the bulk
reference Hamiltonian.

The extended $O(\theta^2 k^2)$ theory also yields some interesting
conclusions regarding operator ordering in the effective
kinetic-energy operator $T$ in effective-mass theory.  Much previous
work has examined various possible ways of choosing the exponents
$\alpha$, $\beta$, and $\gamma$ in the von Roos parametrization
\cite{vRoos83}
\begin{equation}
  T_{\mathrm{vR}} = \tfrac14 (m^{\alpha} p m^{\beta} p m^{\gamma} +
  m^{\gamma} p m^{\beta} p m^{\alpha}) , \label{eq:vonRoos}
\end{equation}
where $p$ is the momentum operator (in one dimension), $m$ is the
effective mass, and $\alpha + \beta + \gamma = -1$.  Morrow and
Brownstein \cite{MoBr84,MoBr85} have argued that only exponents
satisfying $\alpha = \gamma$ are physically permissible in abrupt
heterostructures, which would rule out seemingly reasonable
possibilities such as \cite{Gora69} $T = \tfrac14 ( m^{-1} p^2 + p^2
m^{-1} )$.

The present theory is not consistent with any single operator ordering
of this type.  Instead, the terms of order $p^2$ derived here take the
form of a {\em linear combination} of terms containing
position-dependent functions having {\em all possible} operator
orderings with respect to $p$.  The apparent conflict between this
result and the theory of Morrow and Brownstein \cite{MoBr84,MoBr85} is
resolved by the fact that these are smooth functions of position with
no discontinuity even at an abrupt junction.

The structure of the paper is as follows.  The quadratic-response
theory used in the preceding article \cite{Fore07a} to simplify the
self-consistent pseudopotential is briefly reviewed in Sec.\
\ref{sec:QR_review}.  Section \ref{sec:EFHam} describes the
construction of the envelope-function Hamiltonian, which is also
studied from the perspective of the theory of invariants.  In Sec.\
\ref{sec:mass_parameters}, the parameters of the Hamiltonian are
calculated and discussed for the material systems GaAs/AlAs,
GaAs/Al$_{0.2}$Ga$_{0.8}$As, and In$_{0.53}$Ga$_{0.47}$As/InP\@.  The
valence subband structures for (001) superlattices of these materials
are calculated in Sec.\ \ref{sec:subbands}, which compares the
predictions of various approximate envelope-function models with
``exact'' numerical calculations.  Finally, the results of the paper
are discussed and summarized in Sec.\ \ref{sec:conclusions}.

\section{Quadratic-response theory}

\label{sec:QR_review}

The foundations for this work were laid in the preceding article,
\cite{Fore07a} in which methods are developed for approximating the
self-consistent heterostructure pseudopotential using
quadratic-response theory.  This section presents a brief summary of
the main ideas.

All of the numerical results in this paper are taken from
self-consistent total-energy calculations \cite{Payne92,Mart04}
performed in a plane-wave basis using the \textsc{abinit} soft\-ware
\cite{Gonze02,Gonze05,ABINIT} with nonlocal norm-conserving
pseudopotentials and the local-density approximation (LDA) to
density-functional theory.  Spin-orbit coupling is neglected.  This
model was chosen in order to permit direct comparisons of
envelope-function calculations with ``exact'' numerical calculations
in the relatively large superlattices where envelope-function theory
is valid.  Some technical ingredients of the calculations and the
justification for choosing this particular model are discussed further
in Ref.\ \onlinecite{Fore07a}.  As shown there, the model system
predicts valence-band offsets in reasonably good agreement with
experiment, but (as usual for LDA calculations\cite{Mart04}) does not
predict accurate conduction bands.

The envelope-function Hamiltonian in this paper is not calculated
directly from the ``exact'' superlattice pseudopotential (since that
would require a separate self-consistent calculation for each new
structure), but is instead obtained from the linear and quadratic
response to virtual-crystal perturbations of a bulk reference
crystal. \cite{Fore05b,Fore07a} The reference crystal is defined as
the virtual-crystal average of the bulk constituents (e.g.,
Al$_{0.5}$Ga$_{0.5}$As for GaAs/AlAs and
In$_{0.765}$Ga$_{0.235}$As$_{0.5}$P$_{0.5}$ for
In$_{0.53}$Ga$_{0.47}$As/InP).  The perturbation of the
heterostructure relative to the reference crystal is defined by the
change in pseudopotential
\begin{equation}
  \Delta V_{\mathrm{psp}} (\vect{x}) = \sum_{\alpha} \sum_{\vect{R}}
  \theta^{\alpha}_{\vect{R}} v^{\alpha}_{\mathrm{ion}} (\vect{x} -
  \vect{R}_{\alpha}) ,
\end{equation}
where $v^{\alpha}_{\mathrm{ion}}(\vect{x})$ is the ionic
pseudopotential of atom $\alpha$, $\vect{R}_{\alpha}$ is the position
of atom $\alpha$ in unit cell $\vect{R}$, and
$\theta^{\alpha}_{\vect{R}}$ is the change in fractional weight of
atom $\alpha$ in cell $\vect{R}$ of the heterostructure relative to
the reference crystal.

The fundamental assumption of nonlinear response theory is that the
valence electron density $n$ and the self-consistent pseudopotential
can be expressed as power series in the perturbation
$\theta^{\alpha}_{\vect{R}}$; e.g.,
\begin{subequations} \label{eq:nonlinear}
\begin{align}
  n(\vect{x}) & = n^{(0)}(\vect{x}) + n^{(1)}(\vect{x}) +
  n^{(2)}(\vect{x}) + \cdots , \\ n^{(1)}(\vect{x}) & = \sideset{}{'}
  \sum_{\alpha,\vect{R}} \theta^{\alpha}_{\vect{R}} \Delta
  n^{\alpha}_{\vect{R}} (\vect{x}) , \label{eq:n1} \\
  n^{(2)}(\vect{x}) & = \sideset{}{'} \sum_{\alpha,\vect{R}}
  \sideset{}{'} \sum_{\alpha',\vect{R}'} \theta^{\alpha}_{\vect{R}}
  \theta^{\alpha'}_{\vect{R}'} \Delta
  n^{\alpha\alpha'}_{\vect{R}\vect{R}'} (\vect{x}) . \label{eq:n2}
\end{align}
\end{subequations}
Here $n^{(0)}(\vect{x})$ is the density of the reference crystal,
$n^{(1)}(\vect{x})$ is the total linear response, and
$n^{(2)}(\vect{x})$ is the total quadratic response.  The coefficients
$\Delta n^{\alpha}_{\vect{R}} (\vect{x})$ and $\Delta
n^{\alpha\alpha'}_{\vect{R}\vect{R}'} (\vect{x})$ give the linear and
quadratic response to individual monatomic and diatomic perturbations
of the reference crystal.  The summations in Eqs.\ (\ref{eq:n1}) and
(\ref{eq:n2}) are limited to independent values of
$\alpha$. \cite{Fore05b,Fore07a}

To find all eigenstates of the Hamiltonian, one must know the electron
density $n(\vect{k} + \vect{G})$ for all values of the crystal
momentum $\vect{k}$.  However, the properties of slowly varying
envelope functions depend on the density only in a small neighborhood
of each bulk reciprocal-lattice vector $\vect{G}$.  Within such a
neighborhood, it is convenient to approximate $n(\vect{k} + \vect{G})$
using a power-series expansion.  In a superlattice, the allowed values
of $\vect{k} + \vect{G}$ satisfy $\vect{k} = k \hat{\vect{z}}$ for
small $|\vect{k}|$, where $\hat{\vect{z}}$ is the direction normal to
the interface plane.  The power series for the monatomic and diatomic
response coefficients in Eqs.\ (\ref{eq:n1}) and (\ref{eq:n2})
therefore have the form
\begin{equation}
  \Delta n (\vect{k} + \vect{G}) = \sum_{l=0}^{\infty} (-ik)^l \Delta
  n_l (\vect{G}) . \label{eq:power}
\end{equation}
Similar expansions are used for the local part of the ionic
pseudopotential and the exchange-correlation potential.  In the
present work, the power series for the potentials are approximated
using $0 \le l \le 2$ for the linear potential and $l = 0$ for the
quadratic potential. \cite{Fore05b,Fore07a}

The nonlocal pseudopotential is also expanded in a similar power
series, retaining terms of order $k^4$ in the reference crystal and
order $k^2$ in the heterostructure perturbation (which is linear by
definition).  The methods used to obtain this expansion are described
in Appendix \ref{app:nonlocal}.

\section{Envelope-function Hamiltonian}

\label{sec:EFHam}

The $\vect{k}$-space expansion coefficients for the linear and
quadratic density and short-range potentials may now be used to
construct an envelope-function Hamiltonian. \cite{Fore05b} The entire
set of expansion coefficients including the local and nonlocal
potentials is subjected to the unitary transformation given in Eq.\
(4.15) of Ref.\ \onlinecite{Fore05b}, which converts all matrices from
the plane-wave basis to a Luttinger--Kohn basis of zone-center Bloch
functions for the reference crystal.  Perturbation theory is then used
to reduce these $283 \times 283$ matrices (where 283 is the number of
$\Gamma$ plane waves in the reference-crystal basis\cite{Fore07a}) to
a much smaller dimension by eliminating the coupling between states in
the set $\mathcal{A}$ of interest (e.g., $\mathcal{A} = \{
\Gamma_{15\mathrm{v}} \}$) and all other states.  The Luttinger--Kohn
unitary transformation method
\cite{LuKo55,BirPik74_sec27,Leib75,Leib77,Wink03} is used in order to
obtain energy-independent Hamiltonian parameters.  The basic
perturbation formulas are summarized in Appendix
\ref{app:perturbation}.

The renormalized Hamiltonian for set $\mathcal{A}$ in the reference
crystal is defined by its matrix elements (in the Luttinger--Kohn
basis)
\begin{multline}
  \langle n \vect{k} | \bar{H}^{(0)} | n' \vect{k}' \rangle = (E_n
  \delta_{nn'} + k_{i} \pi^{i}_{nn'} + k_{i} k_{j} D^{ij}_{nn'} \\ +
  k_{i} k_{j} k_{l} C^{ijl}_{nn'} + k_{i} k_{j} k_{l} k_{m}
  Q^{ijlm}_{nn'}) \delta_{\vect{k}\vect{k}'} , \label{eq:H}
\end{multline}
the coefficients of which are defined in Appendix \ref{app:bulk}.
Since the set $\mathcal{A} = \{ \Gamma_{15\mathrm{v}} \}$ will be the
focus of subsequent numerical study, symmetry restrictions on the
kinetic momentum matrix $\pi^{i}_{nn'}$ and the inverse effective-mass
tensor $D^{ij}_{nn'}$ are of interest.  From the symmetry of the
zinc-blende crystal, the matrix $\pi^i$ must have the form
\begin{equation}
  \pi^i = -i R | \epsilon_{ijk} | \{ I_j I_k \} ,
\end{equation}
where $\epsilon_{ijk}$ is the antisymmetric unit tensor, $\{ AB \} =
\frac12 (AB + BA)$, and $I_j$ is the $j$ component of the orbital
angular momentum in a basis $\mathcal{A} = \{ |X\rangle, |Y\rangle,
|Z\rangle \}$ of $p$-like orbitals. \cite{Lutt56} The coefficient $R =
-i \pi_{XY}^z$ is real by time-reversal symmetry, and the hermiticity
of the Hamiltonian requires $R = -R = 0$.  Likewise, the $D^{ij}$
matrix is given by \cite{Lutt56}
\begin{multline}
  D^{ij} = L \delta_{ij} 1 + (M - L) \delta_{ij} I_i^2 - N (1 -
  \delta_{ij}) \{ I_i I_j \} \\ + \tfrac{i}{2} K \epsilon_{ijk} I_k ,
\end{multline}
where 1 represents the $3 \times 3$ unit matrix and the constants $L =
D_{XX}^{xx}$, $M = D_{XX}^{yy}$, $N = D_{XY}^{xy} + D_{XY}^{yx}$, and
$K = D_{XY}^{xy} - D_{XY}^{yx}$ are real.  Here $L$, $M$, and $N$
determine the anisotropic $\Gamma_{15\mathrm{v}}$ effective masses,
while $K$ is the effective Land\'e $g$ factor for the valence
band. \cite{Lutt56}

The renormalized linear-response Hamiltonian for set $\mathcal{A}$ is
given by \cite{Fore05b}
\begin{equation}
  \langle n \vect{k} | \bar{H}^{(1)} | n' \vect{k}' \rangle =
  \sideset{}{'} \sum_{\alpha} \theta_{\alpha} (\vect{k} - \vect{k}')
  H^{\alpha}_{nn'} (\vect{k}, \vect{k}') , \label{eq:H1}
\end{equation}
where $\theta_{\alpha} (\vect{k})$ is the Fourier transform
\cite{Fore05b} of the atomic distribution function
$\theta^{\alpha}_{\vect{R}}$ and
\begin{multline}
  H^{\alpha}_{nn'} (\vect{k}, \vect{k}') = E^{\alpha}_{nn'} +
  \Xi^{\alpha} \delta_{nn'} \delta_{\vect{k}\vect{k}'} + k_i
  \pi^{i\alpha}_{nn'} + \pi^{\alpha i}_{nn'} k_{i}' \\ + k_i k_j
  D^{ij\alpha}_{nn'} + k_i D^{i \alpha j}_{nn'} k_j' + D^{\alpha
    ij}_{nn'} k_i' k_j' . \label{eq:Ha}
\end{multline}
Here $\Xi^{\alpha} = 4 \pi n_2^{\alpha} (\vect{0})$ is the nonanalytic
contribution \cite{Fore07a,note:nonanalytic} from the linear
quadrupole moment of the electron density, which merely shifts the
mean energy by a constant; the remaining terms are defined in Appendix
\ref{app:linear}.  The notation has been modified with respect to
Ref.\ \onlinecite{Fore05b} in order to draw attention to similarities
with the bulk Hamiltonian (\ref{eq:H}) and avoid undue proliferation
of symbols.  It should be noted that the superscripts on the
coefficients in Eq.\ (\ref{eq:Ha}) indicate how the coordinate and
momentum operators are ordered; for example, in the coordinate
representation, the term proportional to $D^{i \alpha j}_{nn'}$ has
the form $D^{i \alpha j}_{nn'} p_i \theta_{\alpha} (\vect{x}) p_j$,
where $\vect{p}$ is the momentum operator. \cite{note:k_to_p}

The symmetry of the coefficients in Eq.\ (\ref{eq:Ha}) is the same as
the symmetry of site $\alpha$ in the reference crystal.  For a
zinc-blende reference crystal, the atomic sites have the same point
group $T_d$ as the reference crystal, but in nonsymmorphic crystals
(e.g., diamond) the site symmetry is lower than the crystal
symmetry. \cite{BirPik74_sec27} Thus, the linear momentum matrix for
$\mathcal{A} = \{ \Gamma_{15\mathrm{v}} \}$ has the same form as in
bulk:
\begin{subequations} \label{eq:pia}
\begin{align}
  \pi^{i\alpha} & = -i R^{\cdot\alpha} | \epsilon_{ijk} | \{ I_j I_k
  \} , \\ \pi^{\alpha i} & = -i R^{\alpha\cdot} | \epsilon_{ijk} | \{
  I_j I_k \} ,
\end{align}
\end{subequations}
where the superscript dots are just placeholders to indicate where
$\vect{p}$ is positioned with respect to $\theta_{\alpha} (\vect{x})$.
As in the bulk case, $R^{\cdot\alpha} = -i \pi^{z\alpha}_{XY}$ and
$R^{\alpha\cdot} = -i \pi^{\alpha z}_{XY}$ are real by time-reversal
symmetry; however, for the linear response, hermiticity [i.e.,
$\pi^{\alpha i}_{nn'} = (\pi^{i\alpha}_{n'n})^*$] requires only that
$R^{\cdot\alpha} = -R^{\alpha\cdot}$.  Therefore, unlike the bulk
case, the linear $R$ coefficients are not required to vanish.  As
discussed in Ref.\ \onlinecite{Fore05b} (using a different notation),
the constant $R^{\cdot\alpha}$ generates a $\delta$-function-like
mixing of the $|X\rangle$ and $|Y\rangle$ valence states at a $(001)$
heterojunction. \cite{IvKaRo96} This mixing is considered in greater
detail below.

The linear $D$ tensor likewise has the same form as in bulk material:
\begin{multline}
  D^{ij\alpha} = L^{\cdot\cdot\alpha} \delta_{ij} 1 +
  (M^{\cdot\cdot\alpha} - L^{\cdot\cdot\alpha}) \delta_{ij} I_i^2 \\ -
  N^{\cdot\cdot\alpha} (1 - \delta_{ij}) \{ I_i I_j \} + \tfrac{i}{2}
  K^{\cdot\cdot\alpha} \epsilon_{ijk} I_k , \label{eq:Dija}
\end{multline}
with similar definitions for $D^{i\alpha j}$ and $D^{\alpha ij}$.  All
of the coefficients $L^{\cdot\cdot\alpha}$, etc., are real by
time-reversal symmetry.  Hermiticity of the Hamiltonian requires that
$D^{\alpha ij}_{nn'} = (D^{ji\alpha}_{n'n})^*$ and $D^{i\alpha
j}_{nn'} = (D^{j\alpha i}_{n'n})^*$, which yields constraints of the
form $K^{\cdot\cdot\alpha} = K^{\alpha\cdot\cdot}$, etc., but does not
require any of the constants in (\ref{eq:Dija}) to be zero.

As discussed in the Introduction, for many heterostructures the energy
gap is not very large in comparison to the band offsets, which means
that the linear approximation for the momentum and mass terms used in
Ref.\ \onlinecite{Fore05b} is not very accurate in a single-band
model.  In an effort to learn more about the limits of single-band
models, the perturbative renormalization of the momentum and mass
terms was extended to terms quadratic in $\theta_{\alpha}$, with the
results given in Appendix \ref{app:quad}.  The resulting contributions
can be written in the form
\begin{multline}
  \langle n \vect{k} | \Delta \bar{H}^{(2)} | n' \vect{k}' \rangle =
  \sideset{}{'} \sum_{\alpha,\beta} \sum_{\vect{k}''} \theta_{\alpha}
  (\vect{k} - \vect{k}'') \theta_{\beta} (\vect{k}'' - \vect{k}') \\
  \times H^{\alpha\beta}_{nn'} (\vect{k}, \vect{k}'', \vect{k}') ,
  \label{eq:H2}
\end{multline}
in which
\begin{multline}
  H^{\alpha\beta}_{nn'} (\vect{k}, \vect{k}'', \vect{k}') =
  E^{\alpha\beta}_{nn'} + k_i \pi^{i\alpha\beta}_{nn'} + k_i''
  \pi^{\alpha i\beta}_{nn'} + \pi^{\alpha\beta i}_{nn'} k_{i}' \\ +
  k_i k_j D^{ij\alpha\beta}_{nn'} + k_i D^{i \alpha\beta j}_{nn'} k_j'
  + D^{\alpha\beta ij}_{nn'} k_i' k_j' \\ + k_i k_j'' D^{i\alpha
  j\beta}_{nn'} + k_i'' D^{\alpha ij\beta}_{nn'} k_j'' + D^{\alpha
  i\beta j}_{nn'} k_i'' k_j' . \label{eq:Hab}
\end{multline}
Here again the superscripts indicate the positioning of the associated
operators, with (for example) the term proportional to $\pi^{\alpha
i\beta}_{nn'}$ having the form $\pi^{\alpha i\beta}_{nn'}
\theta_{\alpha} (\vect{x}) p_i \theta_{\beta} (\vect{x})$ in
coordinate space.  Hermiticity of the Hamiltonian gives constraints
such as $D^{\alpha i\beta j}_{nn'} = (D^{j\beta i\alpha}_{n'n})^*$.
It should be noted that Eqs.\ (\ref{eq:H2}) and (\ref{eq:Hab}) include
only those quadratic contributions arising from perturbative
renormalization; the other terms arising from direct multipole
expansions of the quadratic response have a different form and are
considered in Appendix \ref{app:diatomic}.

For the $\Gamma_{15\mathrm{v}}$ Hamiltonian, the coefficients in Eq.\
(\ref{eq:Hab}) also have $T_d$ symmetry, so they are given by obvious
generalizations of Eqs.\ (\ref{eq:pia}) and (\ref{eq:Dija}).  The
hermiticity constraints on the $R$ coefficients are
$R^{\alpha\beta\cdot} = -R^{\cdot\beta\alpha}$ and
$R^{\alpha\cdot\beta} = -R^{\beta\cdot\alpha}$ (which implies that
$R^{\alpha\cdot\alpha} = 0$), and one can also choose these
coefficients to satisfy $R^{\alpha\beta\cdot} = R^{\beta\alpha\cdot}$
because $\theta_{\alpha}$ and $\theta_{\beta}$ commute.  Likewise, the
$D$ coefficients $L$, $M$, $N$, and $K$ all satisfy constraints of the
form $K^{\alpha\beta\cdot\cdot} = K^{\cdot\cdot\beta\alpha}$,
$K^{\alpha\cdot\beta\cdot} = K^{\cdot\beta\cdot\alpha}$,
$K^{\alpha\cdot\cdot\beta} = K^{\beta\cdot\cdot\alpha}$, and
$K^{\cdot\alpha\beta\cdot} = K^{\cdot\beta\alpha\cdot}$, and one can
choose them to satisfy $K^{\alpha\beta\cdot\cdot} =
K^{\beta\alpha\cdot\cdot}$ too.

In a (001) heterostructure, where $\theta_{\alpha} (\vect{x}) =
\theta_{\alpha}(z)$, the bulk Hamiltonian matrix elements of the form
$L p_z^2$ and $M p_z^2$ are replaced (to order $\theta^2$) by
\begin{multline}
  L p_z^2 \rightarrow L^{(0)} p_z^2 + L^{\cdot\cdot\alpha} (p_z^2
  \theta_{\alpha} + \theta_{\alpha} p_z^2) + L^{\cdot\alpha\cdot} p_z
  \theta_{\alpha} p_z \\ + L^{\cdot\cdot\alpha\beta} (p_z^2
  \theta_{\alpha} \theta_{\beta} + \theta_{\alpha} \theta_{\beta}
  p_z^2) + L^{\cdot\alpha\beta\cdot} p_z \theta_{\alpha}
  \theta_{\beta} p_z \\ + L^{\cdot\alpha\cdot\beta} (p_z
  \theta_{\alpha} p_z \theta_{\beta} + \theta_{\beta} p_z
  \theta_{\alpha} p_z) + L^{\alpha\cdot\cdot\beta} \theta_{\alpha}
  p_z^2 \theta_{\beta} ,
\end{multline}
where summation on $\alpha$ and $\beta$ is implicit.  This is a linear
combination of all but one of the von Roos operators \cite{vRoos83}
defined in Eq.\ (\ref{eq:vonRoos}).  If the renormalization were
extended to cubic order, we would find also a term
\begin{equation}
  L^{\alpha\cdot\beta\cdot\gamma} (\theta_{\alpha} p_z \theta_{\beta}
  p_z \theta_{\gamma} + \theta_{\gamma} p_z \theta_{\beta} p_z
  \theta_{\alpha}) .
\end{equation}
Hence, the present derivation from perturbation theory supports not a
single operator having the form (\ref{eq:vonRoos}) with fixed
exponents, but a linear combination of all possible operator
orderings.  As mentioned in the Introduction, this does not lead to
any mathematical or physical difficulties because $\theta_{\alpha}$ is
a smooth function of $\vect{x}$.  The question of whether any
particular terms in the linear combination might happen to be
negligible is considered in Sec.\ \ref{sec:mass_parameters}.

In a similar fashion, bulk matrix elements of the form $N p_x p_z$ are
replaced in a (001) heterostructure by
\begin{multline}
  N p_x p_z \rightarrow N^{(0)} p_x p_z + (N^{\cdot\alpha\cdot} + 2
  N^{\cdot\cdot\alpha}) p_x \{ p_z, \theta_{\alpha} \} \\ +
  (N^{\cdot\alpha\beta\cdot} + N^{\{ \alpha\cdot\beta\cdot \}} + 2
  N^{\cdot\cdot\alpha\beta}) p_x \{ p_z, \theta_{\alpha}
  \theta_{\beta} \} \\ + (N^{\alpha\cdot\cdot\beta} + N^{\{
  \alpha\cdot\beta\cdot \}}) p_x \theta_{\alpha} p_z \theta_{\beta} ,
\end{multline}
where $N^{\{\alpha\cdot\beta\cdot\}} = \tfrac12
(N^{\alpha\cdot\beta\cdot} + N^{\beta\cdot\alpha\cdot})$.  Note that
the final term proportional to $\theta_{\alpha} p_z \theta_{\beta}$ is
not found in the usual symmetrization recipe for envelope-function
Hamiltonians.  \cite{LinLiuSham85,EpScCo87}

In a (001) heterostructure, the contribution of the $R$ terms to the
Hamiltonian is
\begin{multline}
  H_{R} = -i 2 \{ I_x I_y \} ( R^{\cdot\alpha} [p_z, \theta_{\alpha}]
  + R^{\cdot\alpha\beta} [ p_z, \theta_{\alpha} \theta_{\beta} ] \\ +
  R^{\alpha\cdot\beta} \theta_{\alpha} [ p_z, \theta_{\beta} ] ) ,
  \label{eq:H_R}
\end{multline}
which mixes the $|X\rangle$ and $|Y\rangle$ valence states because
\begin{equation}
    2 \{ I_x I_y \} = \left[
      \begin{array}{rrr}
	0 & -1 & 0 \\
	-1 & 0 & 0 \\
        0 & 0 & 0
      \end{array} \right] .
\end{equation}
The function $\theta_{\alpha}(z)$ is a smooth step-like function,
hence $i [p_z, \theta_{\alpha}] = d \theta_{\alpha} / dz$ is localized
at interfaces and has the form of a macroscopic average of the Dirac
$\delta$ function.  The function $\theta_{\alpha} [ p_z,
\theta_{\beta} ]$ is also localized, but the associated term in
(\ref{eq:H_R}) cannot be written as a simple derivative because
$R^{\alpha\cdot\beta} = -R^{\beta\cdot\alpha}$.  Mixing of the type
(\ref{eq:H_R}) has been studied by Ivchenko {\em et
al}.\cite{IvKaRo96}

The contribution of the $K$ terms in a (001) heterostructure is very
similar:
\begin{multline}
    H_{K} = \frac{i}{2} (p_x I_y - p_y I_x) \bigl(
    K^{\cdot\alpha\cdot} [ p_z, \theta_{\alpha} ] \\ +
    (K^{\cdot\alpha\beta\cdot} + K^{\{\alpha\cdot\beta\cdot\}}) [ p_z,
    \theta_{\alpha} \theta_{\beta} ] \\ + (K^{\alpha\cdot\beta\cdot} -
    K^{\beta\cdot\alpha\cdot}) \theta_{\alpha} [ p_z, \theta_{\beta} ]
    \bigr) . \label{eq:H_K}
\end{multline}
Here the operator $(p_x I_y - p_y I_x)$ is analogous to the Rashba
coupling \cite{Rashba60,Leib77,BychRash84} $(p_x \sigma_y - p_y
\sigma_x)$ in the $\Gamma_6$ conduction band, so contributions of the
form (\ref{eq:H_K}) have been referred to as the valence-band Rashba
coupling. \cite{Wink00,Wink02,Wink03} As discussed in Ref.\
\onlinecite{Fore05b}, this type of coupling was introduced in Ref.\
\onlinecite{Fore93} under an approximation that is equivalent to
assuming that the Rashba coefficient is the same as the effective
Land\'e factor $K$ in bulk material.  This has the advantage of
reducing the number of unknown parameters, since $K$ is known from
magnetoabsorption experiments (see, e.g., Ref.\ \onlinecite{PiBr66}).
However, the bulk Land\'e factor to order $\theta^2$ is given by
\begin{subequations}
  \label{eq:K_bulk}
  \begin{align}
    K & = K^{(0)} + \theta_{\alpha} K^{\alpha} + \theta_{\alpha}
    \theta_{\beta} K^{\alpha\beta} , \\ \intertext{where} K^{\alpha} &
    = K^{\cdot\alpha\cdot} + 2 K^{\cdot\cdot\alpha} , \label{eq:Ka} \\
    K^{\alpha\beta} & = K^{\cdot\alpha\beta\cdot} + 2
    K^{\cdot\cdot\alpha\beta} + K^{\cdot\alpha\cdot\beta} +
    K^{\alpha\cdot\beta\cdot} + K^{\alpha\cdot\cdot\beta} ,
    \label{eq:Kab}
  \end{align}
\end{subequations}
which shows that the Rashba coupling (\ref{eq:H_K}) is generally
independent of the bulk Land\'e factor.  To linear order, replacing
the Rashba coefficient with $K$ would be a good approximation if $2
|K^{\cdot\cdot\alpha}| \ll |K^{\cdot\alpha\cdot}|$.  As shown in Sec.\
\ref{sec:mass_parameters}, this is true in some materials but not in
others; hence, it cannot be presumed to hold true in general.

\section{Effective-mass parameters}

\label{sec:mass_parameters}

In this section, the numerical values of the envelope-function
parameters calculated for the model system are examined to see whether
any general conclusions can be drawn regarding the structure of the
Hamiltonian in Sec.\ \ref{sec:EFHam}.  Values of the Luttinger
parameters \cite{Lutt56} $L$, $M$, $N$, and $K$ calculated for various
bulk materials are listed in Table \ref{table:Lutt_bulk_GaAs}.
\begin{table}
\caption{\label{table:Lutt_bulk_GaAs}Luttinger parameters for several
bulk compounds and their virtual-crystal averages.}
\begin{ruledtabular}
  \begin{tabular}{lddd}
    & \multicolumn{1}{c}{GaAs} &
    \multicolumn{1}{c}{Al$_{0.5}$Ga$_{0.5}$As} &
    \multicolumn{1}{c}{AlAs} \\ \hline
      $L$ & -5.028 & -3.648 & -2.863 \\
      $M$ & -1.511 & -1.326 & -1.126 \\
      $N$ & -6.146 & -4.676 & -3.792 \\
      $K$ & -2.086 & -0.993 & -0.514 \\ \hline
    & \multicolumn{1}{c}{In$_{0.53}$Ga$_{0.47}$As} &
    \multicolumn{1}{c}{In$_{0.765}$Ga$_{0.235}$As$_{0.5}$P$_{0.5}$} &
    \multicolumn{1}{c}{InP} \\ \hline
      $L$ & -5.596 & -4.187 & -3.317 \\
      $M$ & -1.385 & -1.253 & -1.119 \\
      $N$ & -6.590 & -5.071 & -4.087 \\
      $K$ & -2.797 & -1.609 & -0.974
  \end{tabular}
\end{ruledtabular}
\end{table}
This table shows that the parameters are of order 1 (in atomic units),
and that their dependence on material composition is not linear.  For
example, the change (relative to the reference crystal) in $K$ is
about $-1$ for GaAs but only $\frac12$ for AlAs.  It should be noted
that the bulk $K$ values reported here include only the contributions
from \kp\ renormalization, since the asymmetric terms in the nonlocal
pseudopotential \cite{Fore06a} cannot be determined by polynomial
fitting. \cite{note:Knl}

To obtain a measure of the accuracy of the linear and quadratic
approximations, the change in effective-mass parameters for various
bulk compounds relative to the reference crystal was calculated in
these approximations using expressions of the form (\ref{eq:K_bulk})
and compared with the exact changes.  The results of these
calculations are given in Table \ref{table:Lutt_quad}.
\begin{table*}
\caption{\label{table:Lutt_quad}Difference between Luttinger
parameters of bulk crystals and average reference crystal: Comparison
of linear and quadratic approximations with exact differences.  Error
columns give per cent relative error.}
\begin{ruledtabular}
  \begin{tabular}{lw{6}w{4}w{6}w{5}w{6}w{6}w{4}w{6}w{5}w{6}}
    & \multicolumn{5}{c}{GaAs}
    & \multicolumn{5}{c}{Al$_{0.2}$Ga$_{0.8}$As} \\ \hline
    & \multicolumn{2}{c}{Linear}
    & \multicolumn{2}{c}{Quadratic}
    & \multicolumn{1}{c}{Exact}
    & \multicolumn{2}{c}{Linear}
    & \multicolumn{2}{c}{Quadratic}
    & \multicolumn{1}{c}{Exact} \\
    & \multicolumn{1}{c}{Value}
    & \multicolumn{1}{c}{Error}
    & \multicolumn{1}{c}{Value}
    & \multicolumn{1}{c}{Error}
    & \multicolumn{1}{c}{Value}
    & \multicolumn{1}{c}{Value}
    & \multicolumn{1}{c}{Error}
    & \multicolumn{1}{c}{Value}
    & \multicolumn{1}{c}{Error}
    & \multicolumn{1}{c}{Value} \\ \hline
    $\Delta L$ & -0.329710 & -7.22\% & -0.353495 & -0.526\% & -0.355363
               &  0.329710 &  7.21\% &  0.305924 & -0.526\% &  0.307542 \\
    $\Delta M$ & -0.035986 &  0.99\% & -0.035632 & -0.003\% & -0.035634
               &  0.035986 & -0.98\% &  0.036339 & -0.004\% &  0.036341 \\
    $\Delta N$ & -0.347108 & -6.83\% & -0.370680 & -0.501\% & -0.372548
               &  0.347108 &  6.75\% &  0.323536 & -0.498\% &  0.325154 \\
    $\Delta K$ & -0.273357 & -8.70\% & -0.297525 & -0.623\% & -0.299389
               &  0.273357 &  8.99\% &  0.249189 & -0.644\% &  0.250803 \\
    \hline
    & \multicolumn{5}{c}{GaAs}
    & \multicolumn{5}{c}{AlAs} \\ \hline
    $\Delta L$ & -0.9968 & -27.8\% & -1.2706 &  -7.93\% & -1.3800
               &  0.9968 &  27.0\% &  0.7230 &  -7.85\% &  0.7846 \\
    $\Delta M$ & -0.1933 &   4.5\% & -0.1855 &   0.24\% & -0.1850
               &  0.1933 &  -3.6\% &  0.2011 &   0.27\% &  0.2005 \\
    $\Delta N$ & -1.0919 & -25.7\% & -1.3609 &  -7.43\% & -1.4701
               &  1.0919 &  23.5\% &  0.8229 &  -6.93\% &  0.8841 \\
    $\Delta K$ & -0.7004 & -35.9\% & -0.9828 & -10.05\% & -1.0926
               &  0.7004 &  45.9\% &  0.4179 & -12.92\% &  0.4799 \\
    \hline
    & \multicolumn{5}{c}{In$_{0.53}$Ga$_{0.47}$As}
    & \multicolumn{5}{c}{InP} \\ \hline
    $\Delta L$ & -1.1067 & -21.5\% & -1.3650 &  -3.12\% & -1.4091
               &  1.1067 &  27.2\% &  0.8483 &  -2.53\% &  0.8703 \\
    $\Delta M$ & -0.1338 &   2.1\% & -0.1319 &   0.63\% & -0.1311
               &  0.1338 &  -0.6\% &  0.1358 &   0.77\% &  0.1347 \\
    $\Delta N$ & -1.2192 & -19.7\% & -1.4758 &  -2.84\% & -1.5189
               &  1.2192 &  24.0\% &  0.9626 &  -2.13\% &  0.9835 \\
    $\Delta K$ & -0.8778 & -26.1\% & -1.1431 &  -3.77\% & -1.1879
               &  0.8778 &  38.1\% &  0.6125 &  -3.63\% &  0.6355
  \end{tabular}
\end{ruledtabular}
\end{table*}
The top part of the table considers the changes in GaAs and
Al$_{0.2}$Ga$_{0.8}$As relative to the reference crystal
Al$_{0.1}$Ga$_{0.9}$As.  Here the linear approximation is accurate to
better than 10\%, while the quadratic approximation is accurate to
better than 1\%.  Since the linear changes are already a small
perturbation in this case, the linear approximation for
GaAs/Al$_{0.2}$Ga$_{0.8}$As should be adequate for most purposes.

Note that $M$, which determines the mass of heavy holes along the
$\langle 100 \rangle$ directions, is much more accurate than $L$, $N$,
and $K$.  This happens because the coupling of $\Gamma_{15\mathrm{v}}$
to $\Gamma_{1\mathrm{c}}$ does not contribute to $M$, \cite{Kane80} so
the remote bands in $M$ are more remote and the heterostructure
perturbation for $M$ is ``shallower'' than for the other parameters.
As discussed in the Introduction, one can achieve a similar effect for
$L$, $N$, and $K$ by including $\Gamma_{1\mathrm{c}}$ in the set
$\mathcal{A}$. \cite{Fore07c}

For the case of GaAs/AlAs (with Al$_{0.5}$Ga$_{0.5}$As as the
reference crystal), the linear approximation is off by nearly 50\% in
some cases, while the quadratic approximation is accurate to within
13\% for $K$ and to within 8\% for the other parameters.  The
quadratic approximation is more accurate for
In$_{0.53}$Ga$_{0.47}$As/InP, with a maximum error of less than 4\%.
Although these results are not perfect, an accuracy of around 10\% in
a small perturbation is good enough in many cases, and as shown in
Sec.\ \ref{sec:subbands}, the quadratic approximation for $\mathcal{A}
= \{ \Gamma_{15\mathrm{v}} \}$ yields reasonably good band structures
over a limited range of energies (although not as good as for a
multiband Hamiltonian).

Calculated values of various parameters in the $\Gamma_{15\mathrm{v}}$
linear-response Hamiltonian (\ref{eq:Ha}) are listed in Table
\ref{table:mass_hetero}.
\begin{table}
  \caption{\label{table:mass_hetero}Linear parameters in the
           $\Gamma_{15\mathrm{v}}$ Hamiltonian.  Here RC stands for
           reference crystal, and the labels light and heavy holes
           refer to the bulk properties in the $\langle 100 \rangle$
           directions.}
  \begin{ruledtabular}
    \begin{tabular}{llccc}
      RC  & & Al$_{0.5}$Ga$_{0.5}$As &
      \multicolumn{2}{c}{\mbox{In$_{0.765}$Ga$_{0.235}$As$_{0.5}$P$_{0.5}$}} \\
      $\alpha$ & & Ga & As & Ga \\ \hline
      Light hole & $L^{\alpha}$ 
                 & $-1.984$ & $-1.806$ & $-0.847$ \\
                 & $L^{\cdot\alpha\cdot}$ & 
		 $-1.341$ & $-0.586$ & $-1.002$ \\
                 & $L^{\cdot\cdot\alpha}$ &
                 $-0.321$ & $-0.610$ & $+0.077$ \\
      Heavy hole & $M^{\alpha}$
                 & $-0.387$ & $-0.329$ & $+0.130$ \\
                 & $M^{\cdot\alpha\cdot}$ &
                 $-0.039$ & $-0.109$ & $+0.093$ \\
                 & $M^{\cdot\cdot\alpha}$ &
                 $-0.174$ & $-0.110$ & $+0.018$ \\
      $k^2$ mixing & $N^{\alpha}$ &
                 $-2.181$ & $-2.074$ & $-0.771$ \\
                 & $N^{\cdot\alpha\cdot}$ &
                 $-1.542$ & $-0.550$ & $-1.085$ \\
                 & $N^{\cdot\cdot\alpha}$ &
                 $-0.320$ & $-0.762$ & $+0.157$ \\
      Land\'e    & $K^{\alpha}$ & 
                 $-1.399$ & $-1.287$ & $-0.993$ \\
      Rashba     & $K^{\cdot\alpha\cdot}$ & 
                 $-1.372$ & $-0.464$ & $-1.089$ \\
                 & $K^{\alpha\cdot\cdot}$ & 
                 $-0.013$ & $-0.411$ & $+0.048$ \\
      $\delta$ mixing & $R^{\cdot\alpha}$ & 
                  $-0.028$ & $-0.017$ & $-0.038$
    \end{tabular}
  \end{ruledtabular}
\end{table}
The linear changes in the bulk Luttinger parameters are determined by
constants $L^{\alpha}$, $M^{\alpha}$, $N^{\alpha}$, and $K^{\alpha}$
of the form (\ref{eq:Ka}).  As noted below Eq.\ (\ref{eq:H_K}), the
linear contribution to the valence-band Rashba coupling is just
$K^{\cdot\alpha\cdot}$.

Many envelope-function calculations in the literature use the
BenDaniel--Duke operator ordering
\cite{BenDaniel66,Bast88,LinLiuSham85,EpScCo87,Burt92} in which it is
assumed that $|L^{\cdot\alpha\cdot}| \gg 2 |L^{\cdot\cdot\alpha}|$,
$|M^{\cdot\alpha\cdot}| \gg 2 |M^{\cdot\cdot\alpha}|$, and
$|N^{\cdot\alpha\cdot}| \gg 2 |N^{\cdot\cdot\alpha}|$.  Inspection of
Table \ref{table:mass_hetero} shows that this is perhaps a tolerable
approximation in some cases (e.g., light holes in GaAs/AlAs), but it
is a bad approximation in others (e.g., heavy holes in GaAs/AlAs).  It
should be noted that Bastard \cite{Bast88} and Burt \cite{Burt92} both
derive the BenDaniel--Duke ordering using variations of L\"owdin
perturbation theory, \cite{Lowd51} which yields energy-dependent mass
parameters.  This type of perturbation theory cannot be used to draw
conclusions about operator ordering in a Hamiltonian with
energy-independent parameters, since Luttinger--Kohn perturbation
theory is qualitatively different.  A detailed comparison of the two
theories is outside the scope of this paper, but it will be noted that
a direct application of L\"owdin perturbation theory (using a power
series expansion to treat the energy dependence of the denominators)
to the present first-principles calculations yields values of
$L^{\cdot\cdot\alpha}$, $M^{\cdot\cdot\alpha}$, and
$N^{\cdot\cdot\alpha}$ that are smaller than those in Table
\ref{table:mass_hetero}, but still not generally negligible.

As mentioned in Sec.\ \ref{sec:EFHam}, it is also common practice to
estimate the Rashba coupling $K^{\cdot\alpha\cdot}$ by the
approximation \cite{Fore93} $K^{\cdot\alpha\cdot} \simeq K^{\alpha}$,
which amounts to an extension of the BenDaniel--Duke hypothesis to the
antisymmetric terms in the Luttinger Hamiltonian. \cite{Lutt56} Table
\ref{table:mass_hetero} shows that this is a good approximation for
cationic perturbations in GaAs/AlAs and In$_{0.53}$Ga$_{0.47}$As/InP,
but not for anionic perturbations in In$_{0.53}$Ga$_{0.47}$As/InP\@.
Hence, as noted in Ref.\ \onlinecite{Fore05b}, this approximation can
only be relied on in general to produce a rough order-of-magnitude
estimate of the Rashba parameter $K^{\cdot\alpha\cdot}$.

Numerical values for the quadratic renormalization terms in Eq.\
(\ref{eq:Hab}) are listed in Table \ref{table:mass_hetero_quad}.
\begin{table}
  \caption{\label{table:mass_hetero_quad}Quadratic parameters in the
           $\Gamma_{15\mathrm{v}}$ Hamiltonian.  Here
           $\bar{N}^{\cdot\alpha\beta\cdot} =
           N^{\cdot\alpha\beta\cdot} + N^{\{\alpha\cdot\beta\cdot\}}$,
           $\bar{N}^{\alpha\cdot\cdot\beta} =
           N^{\alpha\cdot\cdot\beta} + N^{\{\alpha\cdot\beta\cdot\}}$,
           $N^{\{\alpha\cdot\beta\cdot\}} = \tfrac12
           (N^{\alpha\cdot\beta\cdot} + N^{\beta\cdot\alpha\cdot})$,
           and $N^{[\alpha\cdot\beta\cdot]} =
           N^{\alpha\cdot\beta\cdot} - N^{\beta\cdot\alpha\cdot}$,
           with similar definitions for $K$.}
  \begin{ruledtabular}
    \begin{tabular}{ldddd}
      RC  & \multicolumn{1}{c}{Al$_{0.5}$Ga$_{0.5}$As} &
      \multicolumn{3}{c}{In$_{0.765}$Ga$_{0.235}$As$_{0.5}$P$_{0.5}$} \\
      $(\alpha,\beta)$ & \multicolumn{1}{c}{$(\mathrm{Ga},\mathrm{Ga})$} &
      \multicolumn{1}{c}{$(\mathrm{As},\mathrm{As})$} &
      \multicolumn{1}{c}{$(\mathrm{Ga},\mathrm{Ga})$} & 
      \multicolumn{1}{c}{$(\mathrm{As},\mathrm{Ga})$}
      \\ \hline
      $L^{\alpha\beta}$           & -1.087 & -0.633 & -0.668 & -0.262 \\
      $L^{\cdot\alpha\beta\cdot}$ & -0.718 & -0.640 & -1.118 & -0.111 \\
      $L^{\cdot\cdot\alpha\beta}$ & +0.051 & +0.164 & +0.190 & +0.026 \\
      $L^{\alpha\cdot\cdot\beta}$ & -0.007 & -0.081 & -0.002 & +0.010 \\
      $L^{\alpha\cdot\beta\cdot}$ & -0.232 & -0.120 & +0.036 & -0.235 \\
      $L^{\cdot\alpha\cdot\beta}$ & -0.232 & -0.120 & +0.036 & +0.020 \\
      $M^{\alpha\beta}$           & +0.0310 & +0.0014 & -0.0265 & +0.0129 \\
      $M^{\cdot\alpha\beta\cdot}$ & -0.0040 & -0.0119 & -0.1111 & +0.0009 \\
      $M^{\cdot\cdot\alpha\beta}$ & +0.0301 & +0.0209 & +0.0463 & +0.0065 \\
      $M^{\alpha\cdot\cdot\beta}$ & -0.0004 & -0.0077 & +0.0018 & +0.0050 \\
      $M^{\alpha\cdot\beta\cdot}$ & -0.0124 & -0.0104 & -0.0050 & -0.0066 \\
      $M^{\cdot\alpha\cdot\beta}$ & -0.0124 & -0.0104 & -0.0050 & +0.0007 \\
      $N^{\alpha\beta}$
        & -1.074 & -0.638 & -0.679 & -0.252 \\
      $\bar{N}^{\cdot\alpha\beta\cdot}$ 
        & -0.965 & -0.787 & -1.195 & -0.225 \\
      $N^{\cdot\cdot\alpha\beta}$
        & +0.069 & +0.183 & +0.243 & +0.035 \\
      $\bar{N}^{\alpha\cdot\cdot\beta}$
        & -0.248 & -0.216 & +0.031 & -0.097 \\
      $N^{[\alpha\cdot\beta\cdot]}$
        & 0.0 & 0.0 & 0.0 & -0.271 \\
      $K^{\alpha\beta}$
        & -1.128 & -0.660 & -0.629 & -0.278 \\
      $\bar{K}^{\cdot\alpha\beta\cdot}$
        & -0.933 & -0.746 & -0.966 & -0.217 \\
      $K^{\cdot\cdot\alpha\beta}$
        & +0.009 & +0.133 & +0.151 & +0.019 \\
      $\bar{K}^{\alpha\cdot\cdot\beta}$
        & -0.213 & -0.180 & +0.035 & -0.099 \\
      $K^{[\alpha\cdot\beta\cdot]}$
        & 0.0 & 0.0 & 0.0 & -0.248 \\
      $R^{\cdot\alpha\beta}$
        & -0.0076 & +0.0034 & -0.0185 & -0.0034 \\
      $R^{\alpha\cdot\beta}$
        & 0.0 & 0.0 & 0.0 & -0.0048
    \end{tabular}
  \end{ruledtabular}
\end{table}
The bulk values in this table are defined by expressions of the form
(\ref{eq:Kab}).  It should be noted that the present calculations on
(001) supercells do not provide separate values for the constants
$N^{\{\alpha\cdot\beta\cdot\}}$, $N^{\cdot\alpha\beta\cdot}$, and
$N^{\alpha\cdot\cdot\beta}$, since these terms always appear together
in the sums $N^{\cdot\alpha\beta\cdot} +
N^{\{\alpha\cdot\beta\cdot\}}$ and $N^{\alpha\cdot\cdot\beta} +
N^{\{\alpha\cdot\beta\cdot\}}$ (the same is true for $K$).  Table
\ref{table:mass_hetero_quad} is not discussed here beyond a brief
comment that, although BenDaniel--Duke operator ordering is not a very
good approximation in any case, it is typically better for light holes
than for heavy holes, and better for cation perturbations than for
anion perturbations.  (Of course, since the position-dependent
corrections to the bulk value of $M$ are rather small, heavy-hole
calculations are also less sensitive to the choice of operator
ordering.)

Finally, it should be emphasized that the numbers reported here are
not expected to be accurate.  They are merely intended to provide a
crude picture of some of the qualitative features that would be found
in a more accurate quasiparticle calculation.

\section{Superlattice valence subband structure}

\label{sec:subbands}

\begin{figure}
  \includegraphics[width=8.5cm,clip]{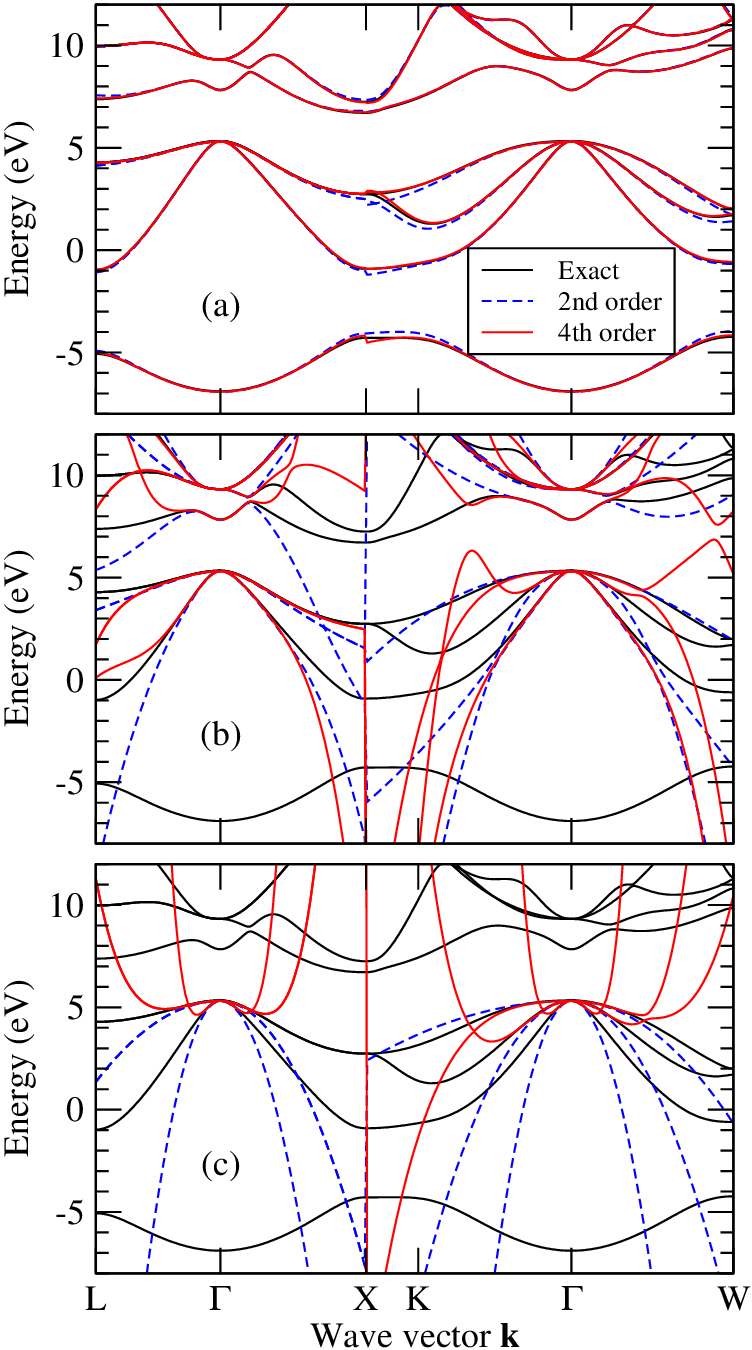}
  \caption{\label{fig:GaAs_bulk_bs} (Color online) Energy band
  structure of bulk Al$_{0.5}$Ga$_{0.5}$As: comparison of exact
  calculation with $O(k^2)$ and $O(k^4)$ $\vect{k} \cdot \vect{p}$
  models.  (a) 283-state $\vect{k} \cdot \vect{p}$ model; (b) 7-state
  $\vect{k} \cdot \vect{p}$ model; (c) 3-state $\vect{k} \cdot
  \vect{p}$ model.}
\end{figure}

In this section, ``exact'' model calculations of the valence subband
structure of (001) superlattices are used to evaluate the accuracy of
various approximate envelope-function models.  As a starting point,
the bulk band structure of Al$_{0.5}$Ga$_{0.5}$As (used as a reference
crystal for GaAs/AlAs) is considered in Fig.\ \ref{fig:GaAs_bulk_bs}.
Part (a) shows the results when all 283 states of the plane-wave basis
are retained in the \kp\ Hamiltonian.  Here it makes little difference
whether the polynomials in the \kp\ Hamiltonian are terminated at
order $k^2$ or $k^4$; both cases provide a good description throughout
the Brillouin zone.  In part (b), the set $\mathcal{A} = \{
\Gamma_{15\mathrm{c}}, \Gamma_{1\mathrm{c}}, \Gamma_{15\mathrm{v}} \}$
contains 7 states (or 14 with spin\cite{Flatte96,PfZa96}).  The
description of the band structure is still accurate near $\Gamma$,
although spurious solutions within the energy gap do occur for both
the $O(k^2)$ and $O(k^4)$ \kp\ models.

Part (c) gives the results for $\mathcal{A} = \{ \Gamma_{15\mathrm{v}}
\}$.  Here there are no spurious solutions in the $O(k^2)$
effective-mass model, but the spurious solutions for the $O(k^4)$
model occur at rather small wave vectors.  The critical point at which
the light-hole band has zero slope is about $\frac16$ of the distance
to the $X$ point.  To prevent problems with spurious solutions, the
envelope functions in a (001) superlattice should contain no Fourier
components beyond this point. \cite{Yang05} Therefore, a superlattice
with a period of 48 monolayers \cite{Fore07a} was chosen as the
standard test case, since this permits the inclusion of 9
envelope-function plane waves within the region $|k_z| \le
\tfrac{4}{48} (4 \pi / a) = \tfrac16 (2 \pi / a)$.  Previous
calculations on empirical pseudopotential models show that this is
sufficient to achieve reasonably accurate results.
\cite{Xia89,WaFrZu97,WaZu97,WaZu99}

The features shown in Fig.\ \ref{fig:GaAs_bulk_bs}(c) have a direct
analog in the Dirac equation for relativistic electrons, for which the
dispersion relation is
\begin{subequations}
  \begin{align}
    E & = \sqrt{p^2 c^2 + m^2 c^4} \\
    & = m c^2 + \frac{p^2}{2m} - \frac{p^4}{8 m^3 c^2} + \cdots .
  \end{align}
\end{subequations}
Here the power series converges only when $|p| < mc$.  If the series
is terminated at order $p^4$, the slope of $E(p)$ changes sign at $p =
\sqrt{2} mc$, which lies outside the region of validity of the power
series.  Although the convergence radius of Luttinger--Kohn
perturbation theory for the \kp\ Hamiltonian is not known, one would
expect it also to be of the same order of magnitude as the critical
point where $E(k)$ changes sign.  Therefore, it is reasonable to
choose this as the cutoff for plane-wave expansions, \cite{Yang05}
even though it may lie slightly outside the convergence radius of the
perturbation power series.

\begin{figure}
  \includegraphics[width=8.5cm,clip]{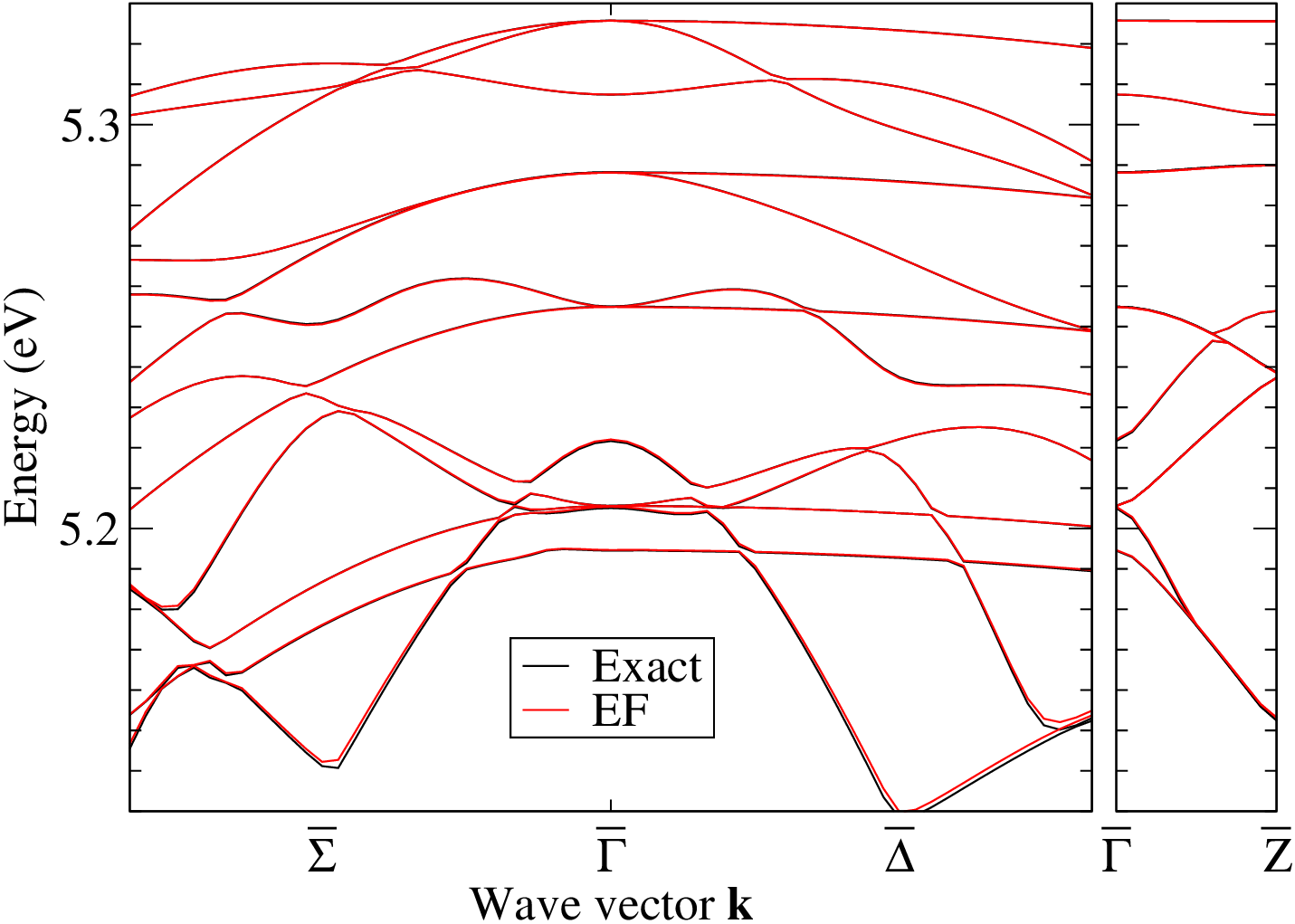}
  \caption{\label{fig:AlGaAs_slat_bs_3} (Color online) Valence band
  structure of a (001) (GaAs)$_{24}$(Al$_{0.2}$Ga$_{0.8}$As)$_{24}$
  superlattice: comparison of exact and 3-state envelope-function (EF)
  calculations.  The EF model is the same as that of Ref.\
  \onlinecite{Fore05b}, and the calculation includes 9 plane waves
  (PW).  Symmetry labels are defined in Refs.\ \onlinecite{ShamLu89}
  and \onlinecite{Koster57}; the $\bar{\Sigma}$ axis corresponds to
  the bulk $\Delta$ axis, and the border of the figure in the
  $\bar{\Sigma}$ direction is 6\% of the distance to the bulk $X$
  point (the border in the $\bar{\Delta}$ or bulk $\Sigma$ direction
  is the same physical distance).}
\end{figure}

The GaAs/Al$_{0.2}$Ga$_{0.8}$As material system historically provided
one of the first direct comparisons between experiment and
effective-mass theory in heterostructures. \cite{Dingle74} Figure
\ref{fig:AlGaAs_slat_bs_3} shows the top 12 valence subbands in a
(001) (GaAs)$_{24}$(Al$_{0.2}$Ga$_{0.8}$As)$_{24}$ superlattice.
The exact model calculations are compared here with a 3-state
$\Gamma_{15\mathrm{v}}$ envelope-function model based on the theory of
Ref.\ \onlinecite{Fore05b}, in which terms of order $k^3$ and $k^4$
are included only for the bulk reference crystal, the mass and
momentum terms are linear in $\theta$, and the potential is quadratic
in $\theta$.  The envelope-function results are in excellent agreement
with the exact calculations; the mean error in each of the first 10
subbands does not exceed 0.1 meV\@.  Note that the valence-band offset
in this system is only 104 meV, which means that the heterostructure
perturbation is indeed shallow \cite{Kohn57b} even in a single-band
$\Gamma_{15\mathrm{v}}$ model.  These results confirm that the theory
of Ref.\ \onlinecite{Fore05b} works very well in the weak-perturbation
limit under which it was derived.

\begin{figure}
  \includegraphics[width=8.5cm,clip]{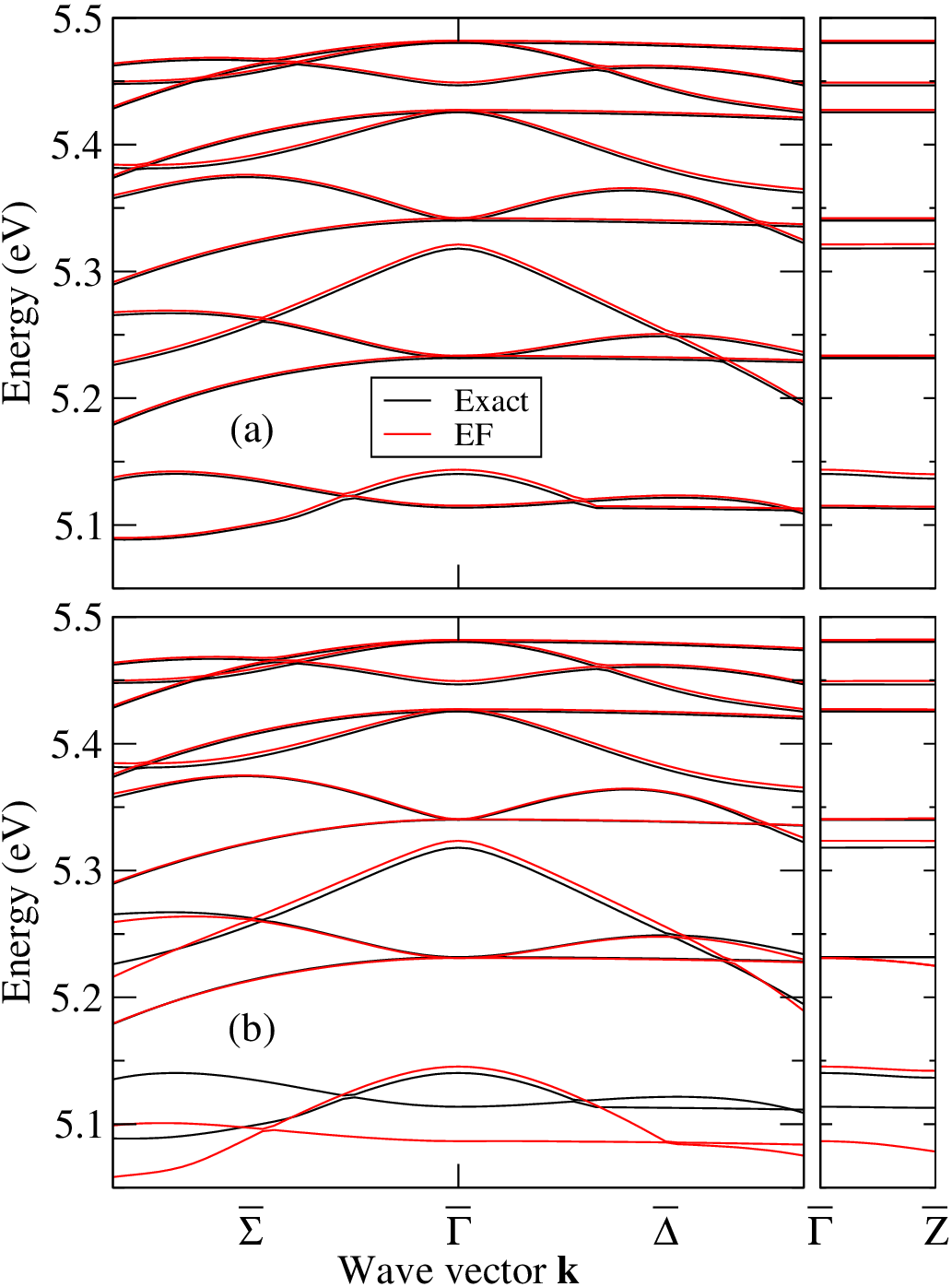}
  \caption{\label{fig:GaAs_slat_bs_283} (Color online) Valence band
  structure of a (001) (GaAs)$_{24}$(AlAs)$_{24}$ superlattice:
  comparison of exact and 283-state EF calculations.  (a) 25 EF plane
  waves; (b) 9 EF plane waves.  Both EF calculations use $O(k^4)$ bulk
  dispersion.}
\end{figure}

The following examples provide a more detailed study of the effects of
various approximations in systems containing stronger perturbations.
Figure \ref{fig:GaAs_slat_bs_283} shows the top 12 valence subbands in
a (001) (GaAs)$_{24}$(AlAs)$_{24}$ superlattice.
The exact model calculations are compared here with a 283-state
envelope-function model that includes all zone-center Bloch functions
explicitly, corresponding to Fig.\ \ref{fig:GaAs_bulk_bs}(a) in the
bulk case.  Figure \ref{fig:GaAs_slat_bs_283}(a) includes 25 plane
waves in the envelope-function model.  The results here are very
accurate, with an error corresponding to about a 2 meV shift that is
nearly the same for all subbands.  (To be more precise, the error in
the ground state is $+1.6$ meV, which is almost the same as the $+1.5$
meV error in the bulk valence-band edge of GaAs calculated in Sec.\
VII~C of the preceding paper.\cite{Fore07a}) Figure
\ref{fig:GaAs_slat_bs_283}(b) shows the effect of reducing the number
of plane waves from 25 to 9.  There is little change for energies
close to the valence-band maximum, but the error in subband 12 is
significant.  Note that the energy range here is much wider than in
Fig.\ \ref{fig:AlGaAs_slat_bs_3}, although it still corresponds
approximately to the valence-band offset. \cite{Fore07a}

\begin{figure}
  \includegraphics[width=8.5cm,clip]{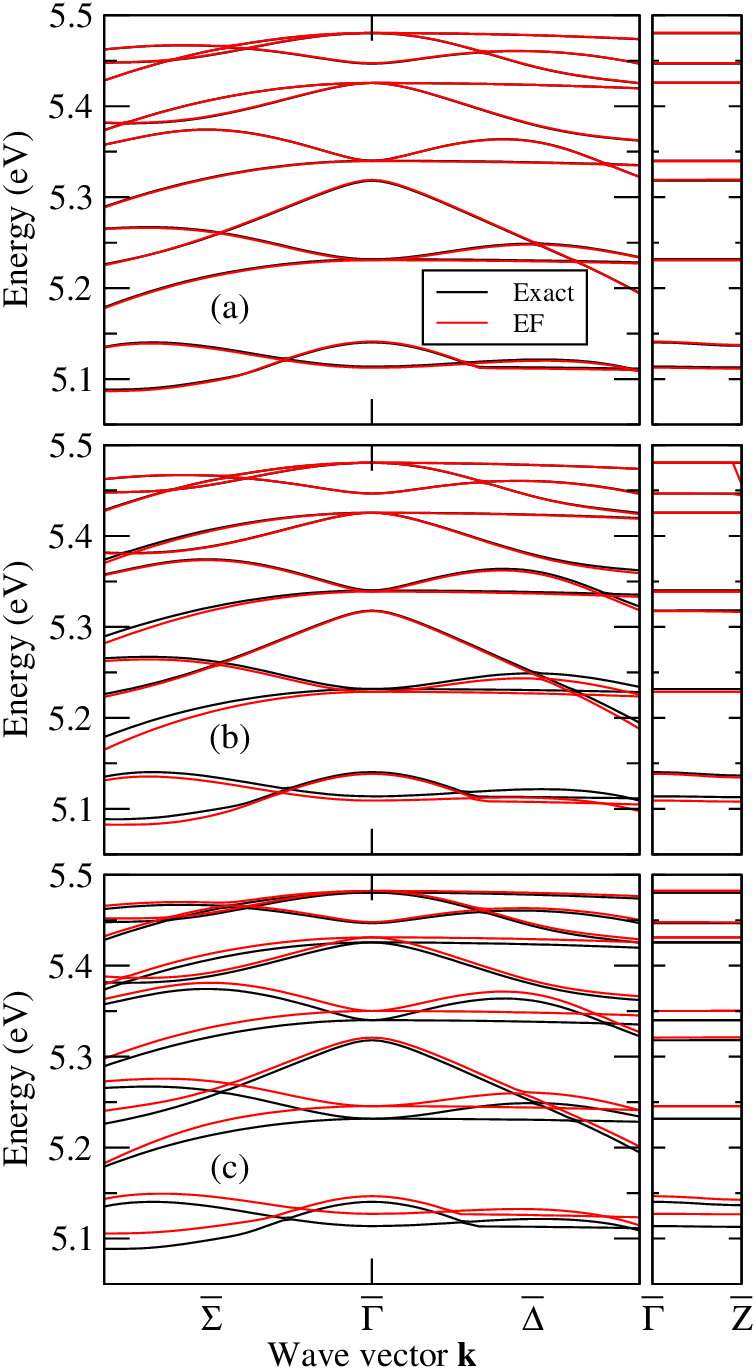}
  \caption{\label{fig:GaAs_slat_bs_7} (Color online) Valence band
  structure of a (001) (GaAs)$_{24}$(AlAs)$_{24}$ superlattice:
  comparison of exact and 7-state EF calculations.  The EF models are
  (a) $O(k^4 \theta^0 + k^2 \theta^1 + k^0 \theta^2)$; (b) $O(k^2
  \theta^1 + k^0 \theta^2)$; (c) $O(k^2 \theta^0 + k^0 \theta^2)$.
  All EF calculations use 25 plane waves and $O(\theta^2)$
  potentials.}
\end{figure}

The effect of reducing set $\mathcal{A}$ to the 7 states in
$\Gamma_{15\mathrm{c}}$, $\Gamma_{1\mathrm{c}}$, and
$\Gamma_{15\mathrm{v}}$ is shown in Fig.\ \ref{fig:GaAs_slat_bs_7}.
Part (a) is just the multiband theory of Ref.\ \onlinecite{Fore05b},
the single-band version of which was used previously in Fig.\
\ref{fig:AlGaAs_slat_bs_3}.  The results here are even slightly better
than in Fig.\ \ref{fig:GaAs_slat_bs_283}(a), which can only be
attributed to a fortuitous cancellation of errors.  In part (b) the
$O(k^4 \theta^0)$ terms are dropped.  The results are still fairly
accurate near the valence-band maximum; however, the peculiar behavior
of (what should be) the ground state near $\bar{Z}$ shows that the
plane-wave cutoff in this case is not quite sufficient to eliminate
all effects of the $O(k^2)$ spurious solutions in Fig.\
\ref{fig:GaAs_bulk_bs}(b).  In Fig.\ \ref{fig:GaAs_slat_bs_7}(c) the
$O(k^2 \theta^1)$ terms are dropped, so that the mass and momentum
matrices are approximated by those of the reference crystal.  Here the
error becomes significant even at fairly small energies, which shows
the importance of including linear terms for multiband models.

\begin{figure}
  \includegraphics[width=8.5cm,clip]{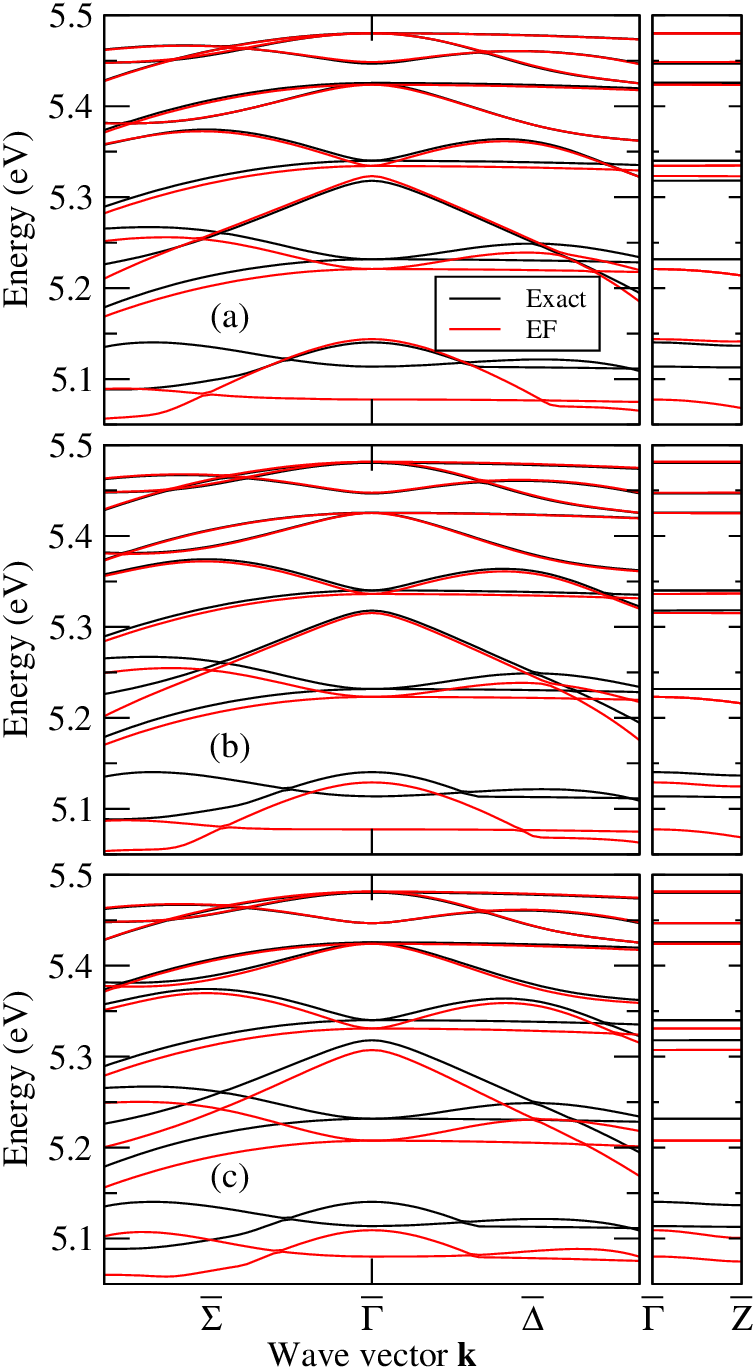}
  \caption{\label{fig:GaAs_slat_bs_3} (Color online) Valence band
  structure of a (001) (GaAs)$_{24}$(AlAs)$_{24}$ superlattice:
  comparison of exact and 3-state EF calculations.  The EF models are
  (a) $O(k^4 \theta^0 + k^2 \theta^1 + k^0 \theta^2)$, 9 PW; (b)
  $O(k^4 \theta^0 + k^2 \theta^2 + k^0 \theta^4)$, 9 PW; (c) $O(k^2
  \theta^2 + k^0 \theta^4)$, 25 PW.}
\end{figure}

\begin{figure}
  \includegraphics[width=8.5cm,clip]{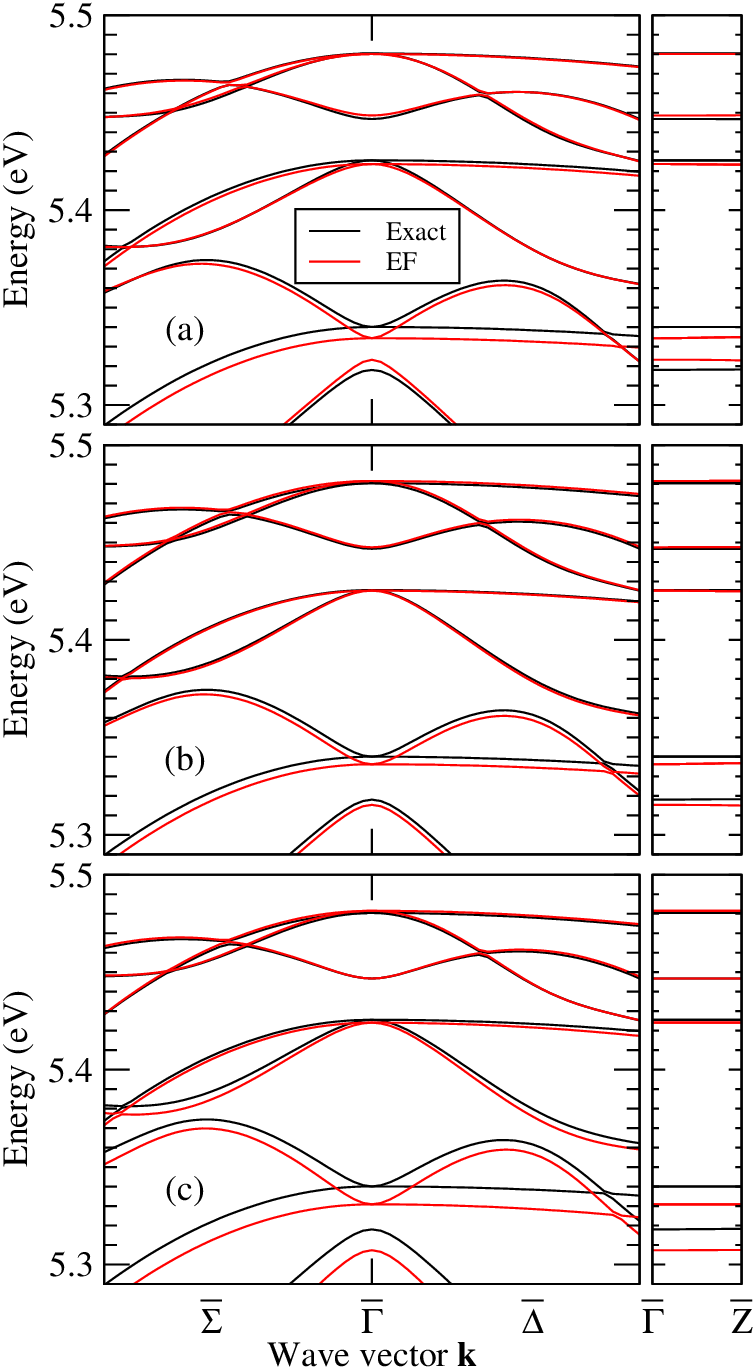}
  \caption{\label{fig:GaAs_slat_bs_3expand} (Color online) Expanded
  view of the top eight valence states from Fig.\
  \ref{fig:GaAs_slat_bs_3}.}
\end{figure}

In Figs.\ \ref{fig:GaAs_slat_bs_3} and \ref{fig:GaAs_slat_bs_3expand},
the set $\mathcal{A}$ is reduced even further to only the three
$\Gamma_{15\mathrm{v}}$ states. %
Figure \ref{fig:GaAs_slat_bs_3} shows the bands on the same energy
scale as before, while Fig.\ \ref{fig:GaAs_slat_bs_3expand} shows an
expanded view of the region near the band edge.  Part (a) uses the
linear mass model of Ref.\ \onlinecite{Fore05b} (the same as in Fig.\
\ref{fig:AlGaAs_slat_bs_3}).  A close inspection of the top three
subbands in Fig.\ \ref{fig:GaAs_slat_bs_3expand}(a) shows evidence of
the errors in the linear mass approximation displayed in Table
\ref{table:Lutt_quad}.  These errors are corrected in part (b), which
includes quadratic corrections to the mass as well as (see Appendix
\ref{app:quad}) terms of order $\theta^4$ in the potential.  This
yields a noticeable improvement, although the quantitative failure for
higher excitations (due in large part to the use of only 9 plane
waves) is still present.  In part (c), the $O(k^4 \theta^0)$ terms are
dropped; since spurious solutions are no longer a problem, the number
of plane waves is increased to 25.  It can be seen that the top three
subbands are still quite accurate under this approximation.

\begin{figure}
  \includegraphics[width=8.5cm,clip]{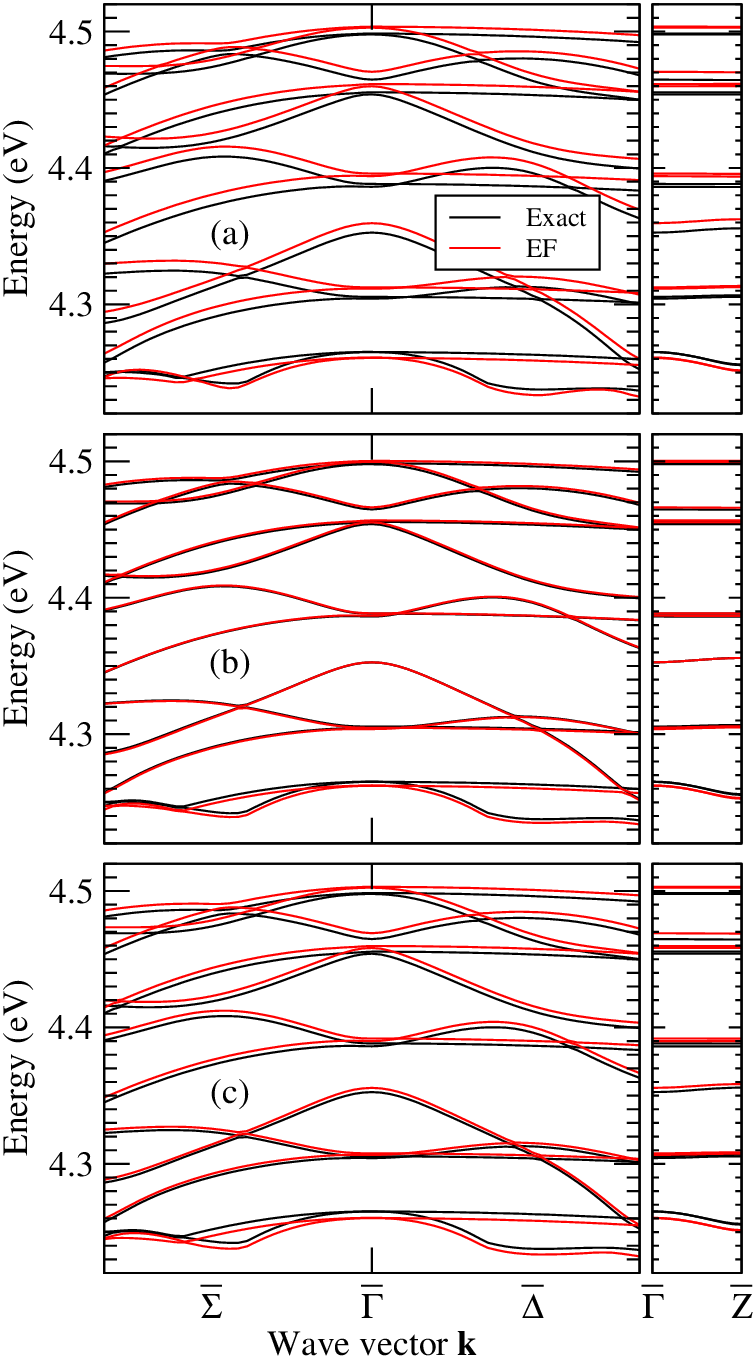}
  \caption{\label{fig:InP_slat_bs_7} (Color online) Valence band
  structure of a (001) (In$_{0.53}$Ga$_{0.47}$As)$_{24}$(InP)$_{24}$
  superlattice: comparison of exact and EF calculations.  (a)
  283-state EF model; (b) 7-state EF model, $O(k^4 \theta^0 + k^2
  \theta^1 + k^0 \theta^2)$; (c) 7-state EF model, $O(k^4 \theta^0 +
  k^2 \theta^2 + k^0 \theta^4)$.  All EF calculations use 25 plane
  waves and $O(k^4)$ bulk dispersion.}
\end{figure}

Figure \ref{fig:InP_slat_bs_7} shows the valence subband structure of
a (001) (In$_{0.53}$Ga$_{0.47}$As)$_{24}$(InP)$_{24}$ superlattice.
Part (a) gives the results obtained from the original 283-state basis.
The error here is larger than in the analogous calculation for
GaAs/AlAs in Fig.\ \ref{fig:GaAs_slat_bs_283}(a); the ground-state
error is $+5.1$ meV, which is close to the error of $+4.5$ meV in the
bulk valence-band edge of In$_{0.53}$Ga$_{0.47}$As calculated in Sec.\
VII~C of Ref.\ \onlinecite{Fore07a}.  The deviation from a constant
shift of about 5 meV is negligible for the first 10 subbands, which
covers the full range of the valence-band offset. \cite{Fore07a}

In Fig.\ \ref{fig:InP_slat_bs_7}(b) the basis is reduced to 7 states
using the linear mass renormalization of Ref.\ \onlinecite{Fore05b}.
The results are quite close to the exact calculation, although again
the improvement over part (a) is fortuitous.  This is demonstrated in
part (c), which includes additional terms of order $\theta^2 k^2$ and
$\theta^4 k^0$.  Most of the bands are shifted slightly upward,
returning nearly to the result from part (a).  The shift is mainly due
to $O(\theta^3 k^0)$ corrections in the renormalized potential [see
Eq.\ (\ref{eq:Eabc})].  Hence, neglecting $O(\theta^3 k^0)$ terms in
perturbative renormalization [Fig.\ \ref{fig:InP_slat_bs_7}(b)]
approximately compensates for the neglect of cubic response terms in
the original Hamiltonian.

Figure 2 of Ref.\ \onlinecite{Fore07c} shows the results of
calculations that are the same as Fig.\ \ref{fig:InP_slat_bs_7}(b),
but for the 4-dimensional set $\mathcal{A} = \{ \Gamma_{1\mathrm{c}},
\Gamma_{15\mathrm{v}} \}$.  The results with $O(\theta^0 k^4)$ terms
are almost identical to Fig.\ \ref{fig:InP_slat_bs_7}(b).  The effect
of dropping the $O(\theta^0 k^4)$ terms is somewhat more significant
than in Fig.\ \ref{fig:GaAs_slat_bs_7}(b), however.

\begin{figure}
  \includegraphics[width=8.5cm,clip]{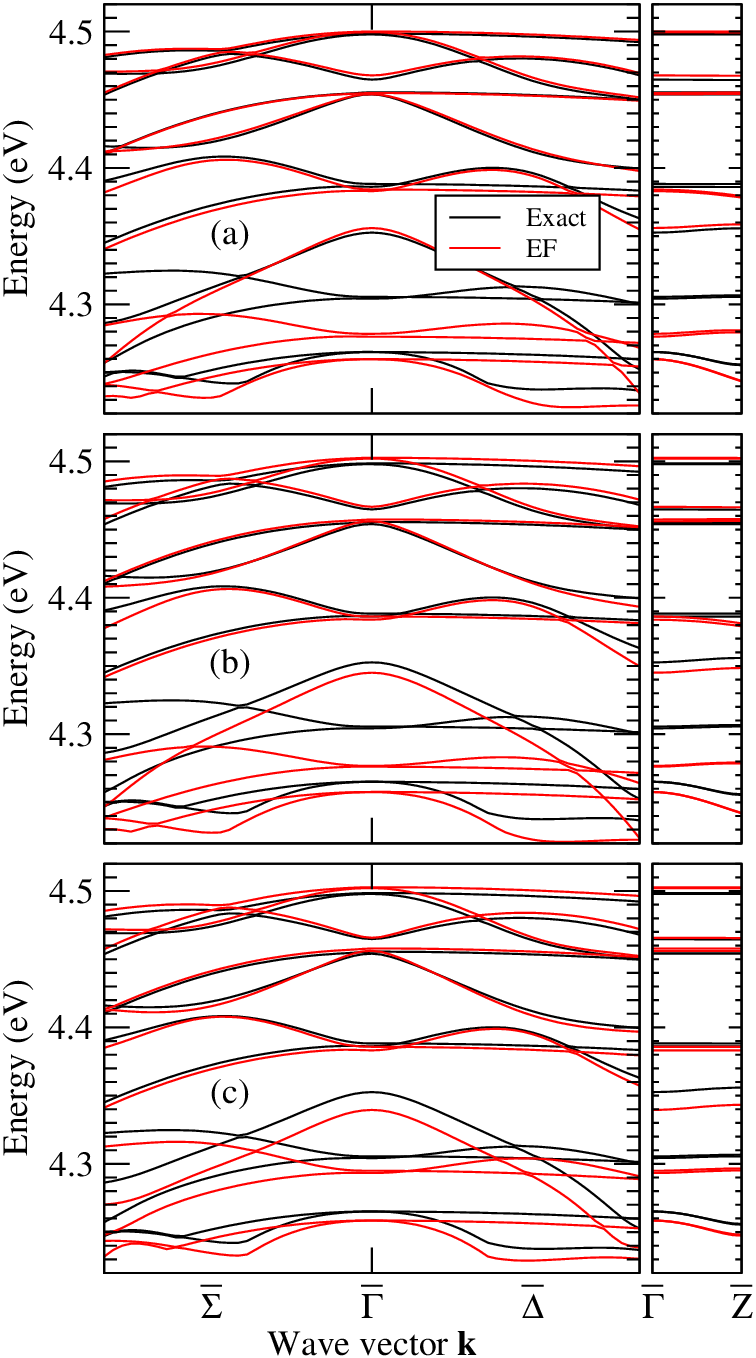}
  \caption{\label{fig:InP_slat_bs_3} (Color online) Valence band
  structure of a (001) (In$_{0.53}$Ga$_{0.47}$As)$_{24}$(InP)$_{24}$
  superlattice: comparison of exact and 3-state EF calculations.  The
  EF models are (a) $O(k^4 \theta^0 + k^2 \theta^1 + k^0 \theta^2)$, 7
  PW; (b) $O(k^4 \theta^0 + k^2 \theta^2 + k^0 \theta^4)$, 7 PW; (c)
  $O(k^2 \theta^2 + k^0 \theta^4)$, 25 PW.}
\end{figure}

Finally, Fig.\ \ref{fig:InP_slat_bs_3} shows the predictions of the
single-band $\Gamma_{15\mathrm{v}}$ model for
In$_{0.53}$Ga$_{0.47}$As/InP\@.
When $O(k^4)$ terms are included in the bulk Hamiltonian, the critical
point of zero slope in the light-hole $\langle 100 \rangle$ dispersion
is closer to $\Gamma$ than it was in GaAs/AlAs, so that now only 7
plane waves can be included in the envelope functions if one wishes to
avoid spurious solutions.  Apart from this restriction, the
envelope-function models used in parts (a), (b), and (c) of Fig.\
\ref{fig:InP_slat_bs_3} are the same as those used in the
corresponding parts of Figs.\ \ref{fig:GaAs_slat_bs_3} and
\ref{fig:GaAs_slat_bs_3expand}.  It can be seen that in this case the
limitations of using only 7 plane waves are sufficiently severe that,
for the first seven subbands, one is better off omitting the $O(k^4)$
terms and including more plane waves.  It should be noted that the
predictions of the single-band Hamiltonian for real
In$_{0.53}$Ga$_{0.47}$As/InP superlattices would likely be
substantially worse than what is shown here (perhaps even
qualitatively incorrect), since the energy gap of
In$_{0.53}$Ga$_{0.47}$As in the model system is 61\% larger than the
experimental value (see Sec.\ IV of the preceding
paper\cite{Fore07a}).

Although it is not visible on the scale of these figures, the double
degeneracy of the $\bar{\Gamma}$ ground state in GaAs/AlAs is removed
in In$_{0.53}$Ga$_{0.47}$As/InP due to the reduction in symmetry from
$D_{2d}$ to $C_{2v}$.  The primary cause of the splitting is mixing of
the $|X\rangle$ and $|Y\rangle$ valence states due to the short-range
interface mixing in Eq.\ (\ref{eq:H_R}) and the long-range interface
dipole potential in Fig.\ 10 of the preceding paper. \cite{Fore07a}
The splitting of the quasidegenerate ground state calculated exactly
and in various envelope-function models is presented in Table
\ref{table:XYsplit}.
\begin{table}
  \caption{\label{table:XYsplit}Splitting of the $\bar{\Gamma}$
           ground-state degeneracy in a (001)
           (In$_{0.53}$Ga$_{0.47}$As)$_{24}$(InP)$_{24}$ superlattice.}
  \begin{ruledtabular}
    \begin{tabular}{lcc} 
      Model & \multicolumn{2}{c}{Diatomic dipole included?} \\
            & Yes & No \\ \hline
      Exact           & 0.639 meV & \\
      EF (283 states) & 0.722 meV & \\
      EF (7 states)   & 0.626 meV & 0.426 meV \\
      EF (3 states)   & 0.585 meV & 
    \end{tabular}
  \end{ruledtabular}
\end{table}
It can be seen that all of the envelope-function models give a
reasonably good estimate of the splitting (which means that they
provide a satisfactory description of the microscopic wave function in
the interface region).  However, when the diatomic dipole terms in
Fig.\ 10 of Ref.\ \onlinecite{Fore07a} (and their associated
polarization of the bulk reference crystal) are omitted, the splitting
of the ground state in the 7-dimensional envelope-function model is
reduced by about one third.  This shows that the practice of fitting
experimental splitting data to short-range interface terms only
\cite{IvKaRo96,Lau02,Szmu04} may give an incorrect description of the
basic physics and overestimate the magnitude of the short-range terms.

\section{Conclusions}

\label{sec:conclusions}

This paper has presented a numerical implementation of the
first-principles envelope-function theory of Ref.\
\onlinecite{Fore05b} in a model system based on superlattice LDA
calculations with norm-conserving pseudopotentials.  The electron
density and potential energy of the superlattice were approximated by
retaining only the linear and quadratic response to the
heterostructure perturbation.  This approximation worked well, with a
net error of about 2 meV in GaAs/AlAs and 5 meV in
In$_{0.53}$Ga$_{0.47}$As/InP\@. \cite{Fore07a} The principal effect of
this error was simply a constant shift of the superlattice energy
eigenvalues.

The density and short-range potentials were then approximated further
using truncated multipole expansions (i.e., power series in $k$),
retaining terms of order $k^2$ in the linear potential and $k^0$ in
the quadratic potential.  This had no effect on the macroscopic
density and potential in bulk, but it generated some additional error
(due primarily to the truncation of the linear density response) in a
narrow region near the interfaces. \cite{Fore07a} This error was
confirmed to be negligible for slowly varying envelope functions.

The approximate Hamiltonian was transformed from the original
plane-wave basis to a Luttinger--Kohn basis using zone-center Bloch
functions of the reference crystal.  A Luttinger--Kohn unitary
transformation was then used to eliminate the \kp\ and
potential-energy coupling between the $\mathcal{A}$ states of interest
and the remote $\mathcal{B}$ states.  The resulting basis is
material-dependent (due to the potential-energy terms) and
approximates the position dependence of the quasi-Bloch functions in
the heterostructure.  The perturbation theory of Ref.\
\onlinecite{Fore05b} was extended to account for quadratic
renormalization of the mass and momentum parameters.

A 7-state $\{ \Gamma_{15\mathrm{v}}, \Gamma_{1\mathrm{c}},
\Gamma_{15\mathrm{c}} \}$ envelope-function model with linear momentum
and mass renormalization was shown to give a very good description of
the $\Gamma_{15}$ valence subband structure of GaAs/AlAs and
In$_{0.53}$Ga$_{0.47}$As/InP (001) superlattices, although the good
results were partly due to a fortuitous cancellation of errors.
Calculations reported elsewhere \cite{Fore07c} show that similar
results for $\Gamma_{15\mathrm{v}}$ can be obtained from a simpler
4-state $\{ \Gamma_{15\mathrm{v}}, \Gamma_{1\mathrm{c}} \}$ model.  A
3-state $\Gamma_{15\mathrm{v}}$ model gave fairly good results over a
more limited energy range (although it probably would not work as well
in real In$_{0.53}$Ga$_{0.47}$As/InP superlattices, since the energy
gap of In$_{0.53}$Ga$_{0.47}$As in the model system was significantly
greater than the experimental value).  The primary limitation of this
single-band model is a conflict between the need for $O(k^4)$ bulk
terms in order to achieve better accuracy in the excited states and
the sometimes rather severe plane-wave cutoff needed to avoid spurious
solutions generated by the $O(k^4)$ terms.  The 3-state model did,
however, give excellent results for GaAs/Al$_{0.2}$Ga$_{0.8}$As, where
the band offset is small enough to satisfy Kohn's definition
($\lesssim 0.1$ eV)\cite{Kohn57b} of a shallow perturbation.

Dipole terms in the quadratic response were found to produce interface
asymmetry and macroscopic electric fields in the no-common-atom
In$_{0.53}$Ga$_{0.47}$As/InP system. \cite{Fore07a} These terms, which
have $C_{2v}$ symmetry, produce a significant fraction of the
splitting of the quasidegenerate ground state in such systems.
Fitting this splitting to only short-range interface $XY$ mixing terms
may therefore overestimate the short-range terms and omit important
physics.

Numerical results for In$_{0.53}$Ga$_{0.47}$As/InP and GaAs/AlAs
indicate that the linear valence-band Rashba coupling parameter is
well approximated by the bulk effective Land\'e factor $K$ for
cationic perturbations, but that there is a wide disparity for anionic
perturbations.  Therefore, using bulk magnetoabsorption measurements
to evaluate interface parameters such as the Rashba coupling
\cite{note:Rod06,Rod06} cannot generally be relied upon to provide
anything better than a rough order-of-magnitude estimate.  Of course,
the particular numbers generated by the present model would likely
change significantly in a more realistic quasiparticle calculation,
but the discrepancy between the Rashba and Land\'e coefficients is
unlikely to vanish.

The operator ordering of the effective-mass terms at a heterojunction
was found to be more complicated than in many previous models.
Instead of having a single von Roos kinetic-energy operator of the
form shown in Eq.\ (\ref{eq:vonRoos}), perturbation theory yields a
linear combination of terms with all possible operator orderings.
Certain terms are larger than others, however.  As shown by Leibler,
\cite{Leib75,Leib77} to linear order only the BenDaniel--Duke operator
\cite{BenDaniel66} $T_{\mathrm{BD}} = \tfrac12 p m^{-1} p$ and the
Gora--Williams operator \cite{Gora69} $T_{\mathrm{GW}} = \tfrac14 (
m^{-1} p^2 + p^2 m^{-1} )$ arise.  In a simple model where the matrix
$E^{\alpha}_{nn'}$ in Eq.\ (\ref{eq:Ha}) is assumed diagonal, the
former arises in third-order perturbation theory from the
position-dependent energies of remote bands in set $\mathcal{B}$,
whereas the latter comes from the position dependence of the bands in
set $\mathcal{A}$ (see Appendix \ref{app:linear}).  The Zhu--Kroemer
operator \cite{ZhKr83} $T_{\mathrm{ZK}} = \tfrac12 m^{-1/2} p^2
m^{-1/2}$ appears as one of several terms in quadratic
renormalization, and the most general von Roos operator does not occur
until cubic order.

Actually, with repeated use of the commutator $[p, f(z)] = -i
(df/dz)$, one can move the momentum operators into any desired
position.  For example, Morrow and Brownstein have shown that, upon
replacing $\alpha \rightarrow \alpha + \epsilon$ and $\gamma
\rightarrow \alpha - \epsilon$ in Eq.\ (\ref{eq:vonRoos}), the von
Roos operator can be rewritten as \cite{MoBr85}
\begin{equation}
  T_{\mathrm{vR}} = T_{\mathrm{H}} - \frac{1}{2m} \left( \frac{d [\ln
  (m^{\epsilon}) ]}{dz} \right)^2 ,
\end{equation}
where $T_{\mathrm{H}} = \frac12 m^{\alpha} p m^{\beta} p m^{\alpha}$
is the Harrison operator \cite{Har61} and the second term has the form
of a potential energy.  But one can continue this process
indefinitely, for example by writing
\begin{equation}
  T_{\mathrm{H}} = T_{\mathrm{BD}} - \frac{d}{dz} \left( \frac{1}{2m}
  \frac{d [\ln (m^{\alpha})]}{dz} \right) + \frac{1}{2m} \left(
  \frac{d [\ln (m^{\alpha})]}{dz} \right)^2 .
\end{equation}
Hence, the operator ordering in the effective kinetic energy is
nothing but an arbitrary convention, as long as one takes care to
retain all of the effective potential-energy terms generated by
changing conventions. \cite{Young89}

The effective kinetic-energy operator given by the perturbation theory
in this paper does have the advantage that the position-dependent
functions $\theta_{\alpha} (\vect{x})$ appearing in it are smooth
step-like functions (although it should be noted that the position of
the step is different for cations and anions).  One could reduce it to
the conventional BenDaniel--Duke form, or any other desired form, if
one were willing to deal with functions having a more complicated
position dependence near the interface.

This suggests that it may be possible---at least at the
phenomenological level---to extend the perturbative scheme described
here to arbitrarily high order in $\theta$. \cite{note:highorder} From
this perspective, as long as (i) the bulk materials of the
heterostructure are accurately described by an effective-mass equation
and (ii) the heterostructure perturbation series eventually converges
at {\em some} finite order, one can rearrange the operators into some
standard order, yielding a standardized Hamiltonian with parameters
that have a complicated but in principle calculable position
dependence.

Of course, this is unlikely to provide a useful first-principles
calculation method unless the series converges at a fairly low order.
Fortunately, the examples given here demonstrate that linear (in the
multiband case) or quadratic (in the single-band case) renormalization
of the momentum and mass parameters is sufficient to achieve good
results in several typical III-V heterostructures.  Given that
linear-response theory \cite{BaReBaPe89} has produced accurate
predictions of valence-band offsets in many other lattice-matched and
lattice-mismatched systems (see the bibliography of Ref.\
\onlinecite{Fore05b}), it is likely that the present envelope-function
theory can be applied successfully in many systems too.

\begin{acknowledgments}
This work was supported by Hong Kong RGC Grant No.\ 600905.
\end{acknowledgments}

\appendix

\section{Nonlocal pseudopotential}

\label{app:nonlocal}

The nonlocal part of the pseudopotential was handled by polynomial
fitting in $\vect{k}$ space.  For the bulk reference crystal, the
entire plane-wave Hamiltonian matrix $H(\vect{k} + \vect{G}, \vect{k}
+ \vect{G}') \equiv H_{\vect{G}\vect{G}'}(\vect{k})$ was evaluated at
57 points near $\vect{k} = \vect{0}$, including $\vect{k} = \vect{0}$
itself and points of the form $\langle 100 \rangle$, $\langle 110
\rangle$, $\langle 111 \rangle$, $\langle 200 \rangle$, and $\langle
210 \rangle \times \Delta$, where $\Delta$ is some specified interval
(usually set equal to half the magnitude of the smallest superlattice
reciprocal-lattice vector).  It is important to choose a set of
fitting points with $O_h$ symmetry so that the fitted coefficients
maintain the $T_d$ symmetry and time-reversal symmetry of the
Hamiltonian.  The Hamiltonian was fitted to a general polynomial of
order $k^4$ with 35 independent coefficients.  (For general $\vect{G}$
and $\vect{G}'$, there is no special symmetry that can be used to
reduce the number of coefficients.)

For the linear response, the nonlocal pseudopotential $V^{\mathrm{nl}}
(\vect{k} + \vect{g}, \vect{k} + \vect{g}')$ may be written as
$V^{\mathrm{nl}}_{\vect{G}\vect{G}'} \bigl( k_x, k_y, k_z + \tfrac12
(\Delta g_z + \Delta g_z'), \Delta g_z - \Delta g_z' \bigr)$, in which
$\vect{g}$ is a reciprocal-lattice vector of the superlattice and
$\Delta \vect{g} \equiv \vect{g} - \vect{G}$ is an integer multiple of
$(2 \pi / Nd) \hat{\vect{z}}$, where $N$ is the number of monolayers
in the superlattice.  The function
$V^{\mathrm{nl}}_{\vect{G}\vect{G}'} (k_1, k_2, k_3, k_4)$ was fitted
to a general polynomial of order $k^2$ with 15 independent
coefficients using a set of 33 points arranged on a cubic grid.  A
larger grid was tested using fitting polynomials of order $k^4$, but
the difference was negligible so only the simpler method was used.
The nonlocal pseudopotential is purely linear, so no fitting of the
quadratic response was necessary.

\section{Perturbation theory}

\label{app:perturbation}

This appendix defines a set of functions that offer a convenient way
to describe operator ordering in fourth-order Luttinger--Kohn
perturbation theory.  These functions are merely an alternative way of
writing the expressions given on p.\ 205 of Winkler's monograph.
\cite{Wink03}

In Luttinger--Kohn perturbation theory, \cite{LuKo55,BirPik74_sec27}
the total Hamiltonian of the system is written as $H = H_0 + H'$,
where $H_0$ has matrix elements $(H_0)_{mm'} = E_m \delta_{mm'}$.  The
states of the unperturbed Hamiltonian $H_0$ are divided into a set
$\mathcal{A}$ containing the states of interest, and a set
$\mathcal{B}$ containing all other states.  It is assumed that the
energies of $\mathcal{A}$ and $\mathcal{B}$ do not overlap.  A unitary
transformation $\bar{H} = e^{-S} H e^S$ is used to eliminate the
coupling between $\mathcal{A}$ and $\mathcal{B}$ to any desired order
in the perturbation $H'$.

The notation used here is defined as follows.  $M_{\mathcal{AB}}$ is
the block of the matrix $M$ that has rows in set $\mathcal{A}$ and
columns in set $\mathcal{B}$.  The matrix $G$ is defined by
\begin{align}
  (G_{\mathcal{AB}})_{nn'} & = (E_{n} - E_{n'})^{-1} , &
  G_{\mathcal{BA}} & = (G_{\mathcal{AB}})^{\mathrm{T}} ,
\end{align}
where T denotes the transpose.  A dot indicates element-by-element
multiplication of congruent matrices:
\begin{equation}
  (A \cdot B)_{nn'} = A_{nn'} B_{nn'} ,
\end{equation}
whereas juxtaposition denotes ordinary matrix multiplication.

The renormalized Hamiltonian $\bar{H}$ for states in $\mathcal{A}$ is
given to fourth order in $H'$ by \cite{Wink03}
\begin{multline}
  \bar{H}_{\mathcal{AA}} = H_{\mathcal{AA}} + P_2(H', H') + P_3(H',
  H', H') \\ + P_4(H', H', H', H') ,
\end{multline}
where the functions $P_2$, $P_3$, and $P_4$ are defined by
\begin{equation}
  P_2(H^1, H^2) = \tfrac12 [ (H^1_{\mathcal{AB}} \cdot
  G_{\mathcal{AB}}) H^2_{\mathcal{BA}} + H^1_{\mathcal{AB}}
  (G_{\mathcal{BA}} \cdot H^2_{\mathcal{BA}}) ] ,
\end{equation}
\begin{multline}
  P_3(H^1, H^2, H^3) = \tfrac12 \{ [ (H^1_{\mathcal{AB}} \cdot
  G_{\mathcal{AB}}) H^2_{\mathcal{BB}} ] \cdot G_{\mathcal{AB}} \}
  H^3_{\mathcal{BA}} \\ + \tfrac12 H^1_{\mathcal{AB}} \{
  G_{\mathcal{BA}} \cdot [ H^2_{\mathcal{BB}} (G_{\mathcal{BA}} \cdot
  H^3_{\mathcal{BA}})] \} \\ - \tfrac12 \{ [ H^1_{\mathcal{AA}} (
  G_{\mathcal{AB}} \cdot H^2_{\mathcal{AB}} ) ] \cdot G_{\mathcal{AB}}
  \} H^3_{\mathcal{BA}} \\ - \tfrac12 H^1_{\mathcal{AB}} \{
  G_{\mathcal{BA}} \cdot [ ( H^2_{\mathcal{BA}} \cdot G_{\mathcal{BA}}
  ) H^3_{\mathcal{AA}} ] \} ,
\end{multline}
\begin{multline}
  P_4(H^1, H^2, H^3, H^4) =
  \\ = \tfrac12 \bigl[ \bigl( H^1_{\mathcal{AA}} \{ G_{\mathcal{AB}}
  \cdot [ H^2_{\mathcal{AA}} ( G_{\mathcal{AB}} \cdot
  H^3_{\mathcal{AB}} ) ] \} \bigr) \cdot G_{\mathcal{AB}} \bigr]
  H^4_{\mathcal{BA}}
  \\ + \tfrac{1}{2} H^1_{\mathcal{AB}} \bigl[ G_{\mathcal{BA}} \cdot
  \bigl( \{ [ ( H^2_{\mathcal{BA}} \cdot G_{\mathcal{BA}} )
  H^3_{\mathcal{AA}} ] \cdot G_{\mathcal{BA}} \} H^4_{\mathcal{AA}}
  \bigr) \bigr]
  \\ - \tfrac{1}{2} H^1_{\mathcal{AB}} \bigl[ G_{\mathcal{BA}} \cdot
  \bigl( \{ [ H^2_{\mathcal{BB}} ( G_{\mathcal{BA}} \cdot
  H^3_{\mathcal{BA}} ) ] \cdot G_{\mathcal{BA}} \} H^4_{\mathcal{AA}}
  \bigr) \bigr]
  \\ - \tfrac{1}{2} H^1_{\mathcal{AB}} \bigl[ G_{\mathcal{BA}} \cdot
  \bigl( H^2_{\mathcal{BB}} \{ G_{\mathcal{BA}} \cdot [ (
  H^3_{\mathcal{BA}} \cdot G_{\mathcal{BA}} ) H^4_{\mathcal{AA}} ] \}
  \bigr) \bigr]
  \\ - \tfrac{1}{2} \bigl[ \bigl( H^1_{\mathcal{AA}} \{
  G_{\mathcal{AB}} \cdot [ ( H^2_{\mathcal{AB}} \cdot G_{\mathcal{AB}}
  ) H^3_{\mathcal{BB}} ] \} \bigr) \cdot G_{\mathcal{AB}} \bigr]
  H^4_{\mathcal{BA}}
  \\ - \tfrac{1}{2} \bigl[ \bigl( \{ [ H^1_{\mathcal{AA}} (
  G_{\mathcal{AB}} \cdot H^2_{\mathcal{AB}} ) ] \cdot G_{\mathcal{AB}}
  \} H^3_{\mathcal{BB}} \bigr) \cdot G_{\mathcal{AB}} \bigr]
  H^4_{\mathcal{BA}}
  \\ - \tfrac{1}{3} \bigl( \{ [ ( H^1_{\mathcal{AB}} \cdot
  G_{\mathcal{AB}} ) H^2_{\mathcal{BA}} ] ( G_{\mathcal{AB}} \cdot
  H^3_{\mathcal{AB}} ) \} \cdot G_{\mathcal{AB}} \bigr)
  H^4_{\mathcal{BA}}
  \\ - \tfrac{1}{3} H^1_{\mathcal{AB}} \bigl( G_{\mathcal{BA}} \cdot
  \{ ( H^2_{\mathcal{BA}} \cdot G_{\mathcal{BA}} ) [
  H^3_{\mathcal{AB}} ( G_{\mathcal{BA}} \cdot H^4_{\mathcal{BA}} ) ]
  \} \bigr)
  \\ - \tfrac{1}{6} \{ [ ( H^1_{\mathcal{AB}} \cdot G_{\mathcal{AB}} )
  ( H^2_{\mathcal{BA}} \cdot G_{\mathcal{BA}} ) H^3_{\mathcal{AB}} ]
  \cdot G_{\mathcal{AB}} \} H^4_{\mathcal{BA}}
  \\ - \tfrac{1}{6} \{ [ H^1_{\mathcal{AB}} ( G_{\mathcal{BA}} \cdot
  H^2_{\mathcal{BA}} ) ( G_{\mathcal{AB}} \cdot H^3_{\mathcal{AB}} ) ]
  \cdot G_{\mathcal{AB}} \} H^4_{\mathcal{BA}}
  \\ - \tfrac{1}{6} H^1_{\mathcal{AB}} \{ G_{\mathcal{BA}} \cdot 
  [ H^2_{\mathcal{BA}} ( G_{\mathcal{AB}} \cdot H^3_{\mathcal{AB}} )
  ( G_{\mathcal{BA}} \cdot H^4_{\mathcal{BA}} ) ] \}
  \\ - \tfrac{1}{6} H^1_{\mathcal{AB}} \{ G_{\mathcal{BA}} \cdot 
  [ ( H^2_{\mathcal{BA}} \cdot G_{\mathcal{BA}} ) ( H^3_{\mathcal{AB}}
  \cdot G_{\mathcal{AB}} ) H^4_{\mathcal{BA}} ] \}
  \\ + \tfrac{1}{24} ( H^1_{\mathcal{AB}} \cdot G_{\mathcal{AB}} )
  ( H^2_{\mathcal{BA}} \cdot G_{\mathcal{BA}} ) ( H^3_{\mathcal{AB}}
  \cdot G_{\mathcal{AB}} ) H^4_{\mathcal{BA}}
  \\ + \tfrac{1}{24} H^1_{\mathcal{AB}} ( G_{\mathcal{BA}} \cdot 
  H^2_{\mathcal{BA}} ) ( G_{\mathcal{AB}} \cdot H^3_{\mathcal{AB}} )
  ( G_{\mathcal{BA}} \cdot H^4_{\mathcal{BA}} )
  \\ + \tfrac{1}{8} ( H^1_{\mathcal{AB}} \cdot G_{\mathcal{AB}} )
  ( H^2_{\mathcal{BA}} \cdot G_{\mathcal{BA}} ) H^3_{\mathcal{AB}}
  ( G_{\mathcal{BA}} \cdot H^4_{\mathcal{BA}} )
  \\ + \tfrac{1}{8} ( H^1_{\mathcal{AB}} \cdot G_{\mathcal{AB}} )
  H^2_{\mathcal{BA}} ( G_{\mathcal{AB}} \cdot H^3_{\mathcal{AB}} )
  ( G_{\mathcal{BA}} \cdot H^4_{\mathcal{BA}} )
  \\ + \tfrac{1}{2} \bigl[ \bigl( \{ [ ( H^1_{\mathcal{AB}} \cdot
  G_{\mathcal{AB}} ) H^2_{\mathcal{BB}} ] \cdot G_{\mathcal{AB}} \}
  H^3_{\mathcal{BB}} \bigr) \cdot G_{\mathcal{AB}} \bigr]
  H^4_{\mathcal{BA}}
  \\ + \tfrac{1}{2} H^1_{\mathcal{AB}} \bigl[ G_{\mathcal{BA}} \cdot
  \bigl( H^2_{\mathcal{BB}} \{ G_{\mathcal{BA}} \cdot [
  H^3_{\mathcal{BB}} ( G_{\mathcal{BA}} \cdot H^4_{\mathcal{BA}} ) ]
  \} \bigr) \bigr] .
\end{multline}
This way of expressing $\bar{H}_{\mathcal{AA}}$ is particularly useful
when $H'$ is a sum of operators that do not commute, and one wishes to
keep track of the order of the various terms.  In the present case,
$H'$ is a sum of \kp\ terms and potential-energy matrix elements.
Examples are given in the appendices below.

\section{Bulk renormalization}

\label{app:bulk}

In terms of the functions defined in Appendix \ref{app:perturbation},
the renormalized coefficients of order $k^2$, $k^3$, and $k^4$ in the
bulk reference Hamiltonian for set $\mathcal{A}$ are given by
\cite{Fore05b}
\begin{equation}
   D^{ij}_{\mathcal{AA}} = \tilde{D}^{ij}_{\mathcal{AA}} +
   P_2(\pi^{i}, \pi^{j}) ,
\end{equation}
\begin{equation}
   C^{ijk}_{\mathcal{AA}} = \tilde{C}^{ijk}_{\mathcal{AA}} +
   P_2(\pi^{i}, \tilde{D}^{jk}) + P_2(\tilde{D}^{ij}, \pi^{k}) + P_3
   (\pi^{i}, \pi^{j}, \pi^{k}) ,
\end{equation}
\begin{multline}
  Q^{ijkl}_{\mathcal{AA}} = \tilde{Q}^{ijkl}_{\mathcal{AA}} + P_4
  (\pi^{i}, \pi^j, \pi^k, \pi^l) \\ + P_3(\tilde{D}^{ij}, \pi^{k},
  \pi^{l}) + P_3 (\pi^{i}, \tilde{D}^{jk}, \pi^{l}) + P_3 (\pi^{i},
  \pi^{j}, \tilde{D}^{kl}) \\ + P_2(\tilde{D}^{ij}, \tilde{D}^{kl}) +
  P_2(\pi^{i}, \tilde{C}^{jkl}) + P_2(\tilde{C}^{ijk}, \pi^{l}) .
\end{multline}
Here a tilde denotes a quantity before renormalization, which is
obtained by fitting the reference Hamiltonian to a polynomial of order
$k^4$ (as described in Appendix \ref{app:nonlocal}).  The tilde is
omitted on $\pi^i$ because it does not change under renormalization.

\section{Linear renormalization}

\label{app:linear}

The terms in the renormalized $\mathcal{A}$ Hamiltonian that are
linear in $\theta_{\alpha}$ are given by \cite{Fore05b}
\begin{align}
  \pi^{i\alpha}_{\mathcal{AA}} & = \tilde{\pi}^{i\alpha}_{\mathcal{AA}}
  + P_2 (\pi^i, E^{\alpha}) , \\
  \pi^{\alpha i}_{\mathcal{AA}} & = \tilde{\pi}^{\alpha i}_{\mathcal{AA}}
  + P_2 (E^{\alpha}, \pi^i) ,
\end{align}
\begin{multline}
  D^{\alpha ij}_{\mathcal{AA}} = \tilde{D}^{\alpha ij}_{\mathcal{AA}}
  + P_2 (\tilde{\pi}^{\alpha i}, \pi^j) + P_2 (E^{\alpha},
  \tilde{D}^{ij}) \\ + P_3 (E^{\alpha}, \pi^i, \pi^j) ,
\end{multline}
\begin{multline}
  D^{i\alpha j}_{\mathcal{AA}} = \tilde{D}^{i \alpha j}_{\mathcal{AA}}
  + P_2 (\tilde{\pi}^{i \alpha}, \pi^j) + P_2 (\pi^i,
  \tilde{\pi}^{\alpha j}) \\ + P_3 (\pi^i, E^{\alpha}, \pi^j) ,
\end{multline}
\begin{multline}
  D^{ij\alpha}_{\mathcal{AA}} = \tilde{D}^{ij \alpha}_{\mathcal{AA}} +
  P_2 (\pi^i, \tilde{\pi}^{j\alpha}) + P_2 (\tilde{D}^{ij},
  E^{\alpha}) \\ + P_3 (\pi^i, \pi^j, E^{\alpha}) .
\end{multline}
These are the same as the expressions given in Appendix C of Ref.\
\onlinecite{Fore05b}, although written in a different notation.  Once
again, a tilde denotes a quantity before renormalization, which is
obtained from a multipole expansion of the linear density and
short-range potentials (Sec.\ \ref{sec:QR_review} and Ref.\
\onlinecite{Fore07a}) and from fitting the linear nonlocal potential
to a polynomial of order $k^2$ (Appendix \ref{app:nonlocal}).  The
tilde is omitted on $E^{\alpha}$ because it does not change under
renormalization.

\section{Quadratic renormalization}

\label{app:quad}

Perturbative renormalization also generates terms that are quadratic
in $\theta_{\alpha}$.  The only term included in the Hamiltonian of
Ref.\ \onlinecite{Fore05b} was the renormalized potential
\begin{equation}
  E^{\alpha\beta}_{\mathcal{AA}} = P_2 (E^{\alpha}, E^{\beta}) .
\end{equation}
Some of the present calculations also include quadratic
renormalization of the momentum matrix:
\begin{equation}
  \pi^{\alpha\beta i}_{\mathcal{AA}} = P_2 (E^{\alpha},
  \tilde{\pi}^{\beta i}) + P_2 (\tilde{E}^{\alpha\beta}, \pi^i) + P_3
  (E^{\alpha}, E^{\beta}, \pi^i) ,
\end{equation}
\begin{equation}
  \pi^{\alpha i\beta}_{\mathcal{AA}} = P_2 (E^{\alpha}, \tilde{\pi}^{i
  \beta}) + P_2 (\tilde{\pi}^{\alpha i}, E^{\beta}) + P_3 (E^{\alpha},
  \pi^i, E^{\beta}) ,
\end{equation}
\begin{equation}
  \pi^{i\alpha\beta}_{\mathcal{AA}} = P_2 (\tilde{\pi}^{i\alpha},
  E^{\beta}) + P_2 (\pi^i, \tilde{E}^{\alpha\beta}) + P_3 (\pi^i,
  E^{\alpha}, E^{\beta}) ,
\end{equation}
and of the effective masses:
\begin{multline}
  D^{\alpha\beta ij}_{\mathcal{AA}} = P_2 (E^{\alpha},
  \tilde{D}^{\beta ij}) + P_2 (\tilde{E}^{\alpha\beta},
  \tilde{D}^{ij}) \\ + P_3 (E^{\alpha}, E^{\beta}, \tilde{D}^{ij}) +
  P_3 (E^{\alpha}, \tilde{\pi}^{\beta i}, \pi^j) + P_3 (
  \tilde{E}^{\alpha\beta}, \pi^i, \pi^j ) \\ + P_4 ( E^{\alpha},
  E^{\beta}, \pi^i, \pi^j) ,
\end{multline}
\begin{multline}
  D^{i\alpha\beta j}_{\mathcal{AA}} = P_2 (\tilde{\pi}^{i \alpha},
  \tilde{\pi}^{\beta j}) + P_3 (\tilde{\pi}^{i \alpha}, E^{\beta},
  \pi^j) \\ + P_3 (\pi^i, E^{\alpha}, \tilde{\pi}^{\beta j}) + P_3
  (\pi^i, \tilde{E}^{\alpha\beta}, \pi^j) \\ + P_4(\pi^i, E^{\alpha},
  E^{\beta}, \pi^j) ,
\end{multline}
\begin{multline}
  D^{ij\alpha\beta}_{\mathcal{AA}} = P_2 (\tilde{D}^{ij\alpha},
  E^{\beta}) + P_2 (\tilde{D}^{ij}, \tilde{E}^{\alpha\beta}) \\ + P_3
  (\tilde{D}^{ij}, E^{\alpha}, E^{\beta}) + P_3 (\pi^i, \tilde{\pi}^{j
  \alpha}, E^{\beta}) + P_3 (\pi^i, \pi^j, \tilde{E}^{\alpha\beta}) \\
  + P_4 (\pi^i, \pi^j, E^{\alpha}, E^{\beta}) ,
\end{multline}
\begin{multline}
  D^{\alpha i\beta j}_{\mathcal{AA}} = P_2 (\tilde{\pi}^{\alpha i},
  \tilde{\pi}^{\beta j}) + P_2 (E^{\alpha}, \tilde{D}^{i\beta j}) \\ +
  P_3 (E^{\alpha}, \pi^i, \tilde{\pi}^{\beta j}) + P_3 (E^{\alpha},
  \tilde{\pi}^{i \beta}, \pi^j) + P_3 (\tilde{\pi}^{\alpha i},
  E^{\beta}, \pi^j) \\ + P_4 ( E^{\alpha}, \pi^i, E^{\beta}, \pi^j) ,
\end{multline}
\begin{multline}
  D^{i\alpha j\beta}_{\mathcal{AA}} = P_2 (\tilde{\pi}^{i\alpha},
  \tilde{\pi}^{j\beta}) + P_2 (\tilde{D}^{i\alpha j}, E^{\beta}) \\ +
  P_3 (\tilde{\pi}^{i\alpha}, \pi^j, E^{\beta}) + P_3 (\pi^i,
  \tilde{\pi}^{\alpha j}, E^{\beta}) + P_3 (\pi^i, E^{\alpha},
  \tilde{\pi}^{j \beta}) \\ + P_4 ( \pi^i, E^{\alpha}, \pi^j,
  E^{\beta}) ,
\end{multline}
\begin{multline}
  D^{\alpha ij\beta}_{\mathcal{AA}} = P_2 (\tilde{\pi}^{\alpha i},
  \tilde{\pi}^{j\beta}) + P_2 (\tilde{D}^{\alpha ij}, E^{\beta}) + P_2
  (E^{\alpha}, \tilde{D}^{ij \beta}) \\ + P_3 (\tilde{\pi}^{\alpha i},
  \pi^j, E^{\beta}) + P_3 (E^{\alpha}, \tilde{D}^{ij}, E^{\beta}) +
  P_3 (E^{\alpha}, \pi^i, \tilde{\pi}^{j \beta}) \\ + P_4 (E^{\alpha},
  \pi^i, \pi^j, E^{\beta}) .
\end{multline}
In these expressions, several approximations are used.  No fitting of
the momentum and mass terms in the original quadratic Hamiltonian was
performed here; consequently, the unrenormalized parts are set to
zero.  Also, the term $\tilde{E}^{\alpha\beta}$ is the mean
unrenormalized diatomic short-range potential summed over all diatomic
perturbations with the given values of $\alpha$ and $\beta$.

Some of the calculations also include renormalized short-range
potential terms of order $\theta^3$ and $\theta^4$, which were
approximated using the expressions
\begin{equation}
  E^{\alpha\beta\gamma}_{\mathcal{AA}} = P_2 (\tilde{E}^{\alpha\beta},
  E^{\gamma}) + P_2 (E^{\alpha}, \tilde{E}^{\beta\gamma}) + P_3
  (E^{\alpha}, E^{\beta}, E^{\gamma}) , \label{eq:Eabc}
\end{equation}
\begin{multline}
  E^{\alpha\beta\gamma\delta}_{\mathcal{AA}} = P_2
  (\tilde{E}^{\alpha\beta}, \tilde{E}^{\gamma\delta}) + P_3
  (E^{\alpha}, E^{\beta}, \tilde{E}^{\gamma\delta}) \\ + P_3
  (E^{\alpha}, \tilde{E}^{\beta\gamma}, E^{\delta}) + P_3
  (\tilde{E}^{\alpha\beta}, E^{\gamma}, E^{\delta}) \\ + P_4
  (E^{\alpha}, E^{\beta}, E^{\gamma}, E^{\delta}) . \label{eq:Eabcd}
\end{multline}
Corrections of the same order arising from the long-range terms in the
Hartree potential were neglected.  For all of the numerical examples
treated here, the $O(\theta^4)$ terms in Eq.\ (\ref{eq:Eabcd}) were
found to be negligible.

It should be emphasized that the results presented here do not provide
a fully consistent perturbation scheme according to the criteria given
by Takhtamirov and Volkov, \cite{TakVol99b} in which the mean kinetic
energy of the states of interest is assumed to be comparable to the
heterostructure potential-energy perturbation.  According to this
scheme, if one includes the $O(k^2 \theta^2)$ and $O(k^0 \theta^3)$
terms shown here, one should also include terms of order $k^4
\theta^1$ and $k^6 \theta^0$.

However, since these require the use of sixth-order perturbation
theory, such terms were judged to be not worth the effort in a
preliminary investigation of this nature.  Therefore, the results
obtained by adding only the $O(k^2 \theta^2)$ and $O(k^0 \theta^3)$
terms are not expected to be valid for kinetic energies covering the
full range of the band offset, but only for kinetic energies small in
comparison to the band offset.  This is indeed what was found in the
numerical calculations of Sec.\ \ref{sec:subbands}\@.  Likewise, the
$O(k^4 \theta^0)$ terms were found to be less important for states of
small kinetic energy.  These results are merely a reflection of the
fact that a wide quantum well, unlike a hydrogenic impurity,
\cite{BirPik74_sec27} does have states in which the mean kinetic
energy is small compared to the mean perturbing potential.  Thus, in
this sense the theory of low-energy excitations of wide quantum wells
is actually simpler than the corresponding theory of shallow
impurities, because terms of high order in $k$ are of lesser
importance.

\section{Quadratic response}

\label{app:diatomic}

The method used here to handle the quadratic response differs slightly
from that of Ref.\ \onlinecite{Fore05b}.  The quadratic potential
response is given by Eq.\ (3.14) of Ref.\ \onlinecite{Fore05b}:
\begin{equation}
  V^{(2)}(\vect{x}, \vect{x}') = \sideset{}{'} \sum_{\alpha,\vect{R}}
  \sideset{}{'} \sum_{\alpha',\vect{R}'} \theta^{\alpha}_{\vect{R}}
  \theta^{\alpha'}_{\vect{R}'} \Delta
  v^{\alpha\alpha'}_{\vect{R}\vect{R}'} (\vect{x}, \vect{x}') ,
  \label{eq:V2}
\end{equation}
which has the same form as the quadratic density response in Eq.\
(\ref{eq:n2}).  The translation symmetry of the reference crystal
allows $\Delta v^{\alpha\alpha'}_{\vect{R}\vect{R}'} (\vect{x},
\vect{x}')$ to be written as
\begin{equation}
  \Delta v^{\alpha\alpha'}_{\vect{R}\vect{R}'} (\vect{x}, \vect{x}')
  \equiv \Delta v^{\alpha\alpha',\vect{R}' - \vect{R}} (\vect{x} -
  \bar{\vect{R}}_{\alpha\alpha'}, \vect{x}' -
  \bar{\vect{R}}_{\alpha\alpha'}) , \label{eq:dv2}
\end{equation}
in which $\bar{\vect{R}}_{\alpha\alpha'}$ is the midpoint of the two
atoms:
\begin{equation}
  \bar{\vect{R}}_{\alpha\alpha'} = \tfrac12 (\vect{R} + \vect{R}' +
  \bm{\tau}_{\alpha} + \bm{\tau}_{\alpha'}) . \label{eq:Rmid}
\end{equation}
In Eq.\ (\ref{eq:dv2}), the coordinate reference is taken to be
$\bar{\vect{R}}_{\alpha\alpha'}$, whereas in Eq.\ (3.17) of Ref.\
\onlinecite{Fore05b}, it was chosen to be $\vect{R}_{\alpha} \equiv
\vect{R} + \bm{\tau}_{\alpha}$.  The Fourier transform of Eq.\
(\ref{eq:V2}) is \cite{Fore05b}
\begin{equation}
  V^{(2)}(\vect{k}, \vect{k}') = N \sideset{}{'} \sum_{\alpha,
  \alpha', \vect{R}'} \theta^{\alpha \alpha' \vect{R}' } (\vect{k} -
  \vect{k}') \Delta v^{\alpha \alpha' \vect{R}' } (\vect{k},
  \vect{k}') , \label{eq:V2kk}
\end{equation}
where
\begin{equation}
  \theta^{\alpha \alpha' \vect{R}'} (\vect{k}) = \frac{1}{N}
  \sum_{\vect{R}} \theta^{\alpha}_{\vect{R}}
  \theta^{\alpha'}_{\vect{R} + \vect{R}'} e^{-i \vect{k} \cdot [
  \vect{R}_{\alpha} + (\vect{R}' + \bm{\tau}_{\alpha'} -
  \bm{\tau}_{\alpha}) / 2 ]} .
\end{equation}
The coordinate reference (\ref{eq:Rmid}) is arbitrary, but it is
sometimes more convenient for analyzing symmetry properties than the
choice used in Ref.\ \onlinecite{Fore05b}.

In the LDA model used here, the quadratic potential is local [i.e.,
$\Delta v^{\alpha \alpha' \vect{R}' } (\vect{k}, \vect{k}') = \Delta
v^{\alpha \alpha' \vect{R}' } (\vect{k} - \vect{k}')$] since the ionic
pseudopotential is purely linear.  The diatomic potential $\Delta
v^{\alpha \alpha' \vect{R}'} (\vect{q})$ was approximated for small
$\vect{q}$ by keeping only the dipole and quadrupole terms in the
electron density and the $l = 0$ term in the power-series expansion
(\ref{eq:power}) of the short-range potential (which in the quadratic
case consists only of the exchange-correlation potential).  For
$\vect{q}$ near $\vect{G} \ne \vect{0}$, only the $l = 0$ terms were
retained in both the density and short-range potential; see Sec.\ VI~B
of the preceding paper \cite{Fore07a} for further details.


\begin{thebibliography}{61}
\bibitem[{not({\natexlab{a}})}]{note:nonanalytic}
\bibinfo{note}{If one considers single atomic perturbations rather than planar
  perturbations, there are additional nonanalytic contributions from the linear
  octopole and hexadecapole moments of the electron density.\cite{Fore05b} For
  the present case of planar perturbations, these are simply absorbed into the
  definition of the short-range potential using Eq.\ (12) of Ref.\
  \onlinecite{Fore07a}.}
\bibitem[{not({\natexlab{c}})}]{note:Knl}
\bibinfo{note}{They could be obtained from analytical derivatives of the
  nonlocal pseudopotential in $\vect{k}$ space, but that was not done here
  since the primary focus of this paper is the physics of abrupt
  heterojunctions. The value of $K$ also depends on the choice of coupling
  between the nonlocal potential and the magnetic field.\cite{Fore06a}}
\bibitem[{not({\natexlab{d}})}]{note:Rod06}
\bibinfo{note}{Rodina and Alekseev \cite{Rod06} have suggested dividing the
  heterostructure Hamiltonian into a bulk part and a surface part, in which the
  valence-band Rashba parameters for the bulk part are defined to be the same
  as the magnetic Luttinger parameters. However, since by symmetry the surface
  part contains terms with the same mathematical structure as the bulk part,
  this division does not carry any meaning unless one can make some definite
  statement about the magnitude of the surface part. As shown by the model
  calculations presented here, the surface part in this scheme is not generally
  negligible.}
\bibitem[{not({\natexlab{e}})}]{note:highorder}
\bibinfo{note}{This scheme would be limited to the low-energy excitations of a
  wide quantum well or superlattice, as discussed in the remarks at the end of
  Appendix \ref{app:quad}.}
\end{thebibliography}

\end{document}